\documentclass[aps,prl,twocolumn,superscriptaddress,showpacs,preprintnumbers,amsmath,amssymb,tightenlines,floatfix]{revtex4-1}

\usepackage{graphicx} 
\usepackage{dcolumn}  
\usepackage{siunitx}  

\newcommand{\Bdecay}{\bar{B} \to D^{(\ast)} \tau^- \bar{\nu}_\tau}
\newcommand{\BDdecay}{\bar{B} \to D \tau^- \bar{\nu}_\tau}
\newcommand{\BDSdecay}{\bar{B} \to D^{\ast} \tau^- \bar{\nu}_\tau}
\newcommand{\Bdecaynorm}{\bar{B} \to D^{(\ast)} \ell^- \bar{\nu}_\ell}
\newcommand{\BDdecaynorm}{\bar{B} \to D \ell^- \bar{\nu}_\ell}
\newcommand{\BDSdecaynorm}{\bar{B} \to D^{\ast} \ell^- \bar{\nu}_\ell}
\newcommand{\mbc}[1]{M_{{\rm bc}#1}}

\newcommand{\ecl}{E_{\textrm{ECL}}}

\newcommand{\MM}{M_\mathrm{miss}^2}
\newcommand{\MMR}{M_{\text{miss,no }\pi^0}^{2}}
\newcommand{\PLEP}{p_\ell^\ast}
\newcommand{\QSQ}{q^{2}}
\newcommand{\NBO}{o_{\text{NB}}}
\newcommand{\NBTR}{o_{\text{NB}}^\prime}

\DeclareSIPostPower\tothefourth{4}
\DeclareSIUnit\clight{$c$}
\DeclareSIUnit\stddev{$\sigma$}
\newcommand{\unitmass}{\giga\electronvolt/\clight\squared}
\newcommand{\unitmasssquared}{\giga\electronvolt\squared/\clight\tothefourth}

\newcommand{\unitlowenergy}{\mega\electronvolt}

\newcommand{\unitmomentum}{\giga\electronvolt/\clight}
\newcommand{\unitmomentumsquared}{\giga\electronvolt\squared/\clight\squared}
\newcommand{\unitlowmomentum}{\mega\electronvolt/\clight}

\newcommand{\component}[1]{#1}

\graphicspath{{ps}}

\begin{document}

\title{Measurement of the branching ratio of \boldmath$\Bdecay$ relative to \boldmath$\Bdecaynorm$ decays with hadronic tagging at Belle}

\noaffiliation
\affiliation{University of the Basque Country UPV/EHU, 48080 Bilbao}
\affiliation{Beihang University, Beijing 100191}
\affiliation{University of Bonn, 53115 Bonn}
\affiliation{Budker Institute of Nuclear Physics SB RAS, Novosibirsk 630090}
\affiliation{Faculty of Mathematics and Physics, Charles University, 121 16 Prague}
\affiliation{Chonnam National University, Kwangju 660-701}
\affiliation{University of Cincinnati, Cincinnati, Ohio 45221}
\affiliation{Deutsches Elektronen--Synchrotron, 22607 Hamburg}
\affiliation{Justus-Liebig-Universit\"at Gie\ss{}en, 35392 Gie\ss{}en}
\affiliation{Gifu University, Gifu 501-1193}
\affiliation{II. Physikalisches Institut, Georg-August-Universit\"at G\"ottingen, 37073 G\"ottingen}
\affiliation{SOKENDAI (The Graduate University for Advanced Studies), Hayama 240-0193}
\affiliation{Hanyang University, Seoul 133-791}
\affiliation{University of Hawaii, Honolulu, Hawaii 96822}
\affiliation{High Energy Accelerator Research Organization (KEK), Tsukuba 305-0801}
\affiliation{IKERBASQUE, Basque Foundation for Science, 48013 Bilbao}
\affiliation{Indian Institute of Technology Guwahati, Assam 781039}
\affiliation{Indian Institute of Technology Madras, Chennai 600036}
\affiliation{Institute of High Energy Physics, Chinese Academy of Sciences, Beijing 100049}
\affiliation{Institute of High Energy Physics, Vienna 1050}
\affiliation{Institute for High Energy Physics, Protvino 142281}
\affiliation{INFN - Sezione di Torino, 10125 Torino}
\affiliation{Institute for Theoretical and Experimental Physics, Moscow 117218}
\affiliation{J. Stefan Institute, 1000 Ljubljana}
\affiliation{Kanagawa University, Yokohama 221-8686}
\affiliation{Institut f\"ur Experimentelle Kernphysik, Karlsruher Institut f\"ur Technologie, 76131 Karlsruhe}
\affiliation{Kennesaw State University, Kennesaw GA 30144}
\affiliation{King Abdulaziz City for Science and Technology, Riyadh 11442}
\affiliation{Department of Physics, Faculty of Science, King Abdulaziz University, Jeddah 21589}
\affiliation{Korea Institute of Science and Technology Information, Daejeon 305-806}
\affiliation{Korea University, Seoul 136-713}
\affiliation{Kyungpook National University, Daegu 702-701}
\affiliation{\'Ecole Polytechnique F\'ed\'erale de Lausanne (EPFL), Lausanne 1015}
\affiliation{Faculty of Mathematics and Physics, University of Ljubljana, 1000 Ljubljana}
\affiliation{Ludwig Maximilians University, 80539 Munich}
\affiliation{Luther College, Decorah, Iowa 52101}
\affiliation{University of Maribor, 2000 Maribor}
\affiliation{Max-Planck-Institut f\"ur Physik, 80805 M\"unchen}
\affiliation{School of Physics, University of Melbourne, Victoria 3010}
\affiliation{Moscow Physical Engineering Institute, Moscow 115409}
\affiliation{Moscow Institute of Physics and Technology, Moscow Region 141700}
\affiliation{Graduate School of Science, Nagoya University, Nagoya 464-8602}
\affiliation{Kobayashi-Maskawa Institute, Nagoya University, Nagoya 464-8602}
\affiliation{Nara Women's University, Nara 630-8506}
\affiliation{National Central University, Chung-li 32054}
\affiliation{National United University, Miao Li 36003}
\affiliation{Department of Physics, National Taiwan University, Taipei 10617}
\affiliation{H. Niewodniczanski Institute of Nuclear Physics, Krakow 31-342}
\affiliation{Niigata University, Niigata 950-2181}
\affiliation{University of Nova Gorica, 5000 Nova Gorica}
\affiliation{Novosibirsk State University, Novosibirsk 630090}
\affiliation{Osaka City University, Osaka 558-8585}
\affiliation{Pacific Northwest National Laboratory, Richland, Washington 99352}
\affiliation{Peking University, Beijing 100871}
\affiliation{University of Pittsburgh, Pittsburgh, Pennsylvania 15260}
\affiliation{University of Science and Technology of China, Hefei 230026}
\affiliation{Soongsil University, Seoul 156-743}
\affiliation{University of South Carolina, Columbia, South Carolina 29208}
\affiliation{Sungkyunkwan University, Suwon 440-746}
\affiliation{School of Physics, University of Sydney, NSW 2006}
\affiliation{Department of Physics, Faculty of Science, University of Tabuk, Tabuk 71451}
\affiliation{Tata Institute of Fundamental Research, Mumbai 400005}
\affiliation{Excellence Cluster Universe, Technische Universit\"at M\"unchen, 85748 Garching}
\affiliation{Toho University, Funabashi 274-8510}
\affiliation{Tohoku University, Sendai 980-8578}
\affiliation{Earthquake Research Institute, University of Tokyo, Tokyo 113-0032}
\affiliation{Department of Physics, University of Tokyo, Tokyo 113-0033}
\affiliation{Tokyo Institute of Technology, Tokyo 152-8550}
\affiliation{Tokyo Metropolitan University, Tokyo 192-0397}
\affiliation{University of Torino, 10124 Torino}
\affiliation{Utkal University, Bhubaneswar 751004}
\affiliation{CNP, Virginia Polytechnic Institute and State University, Blacksburg, Virginia 24061}
\affiliation{Wayne State University, Detroit, Michigan 48202}
\affiliation{Yamagata University, Yamagata 990-8560}
\affiliation{Yonsei University, Seoul 120-749}
  \author{M.~Huschle}\affiliation{Institut f\"ur Experimentelle Kernphysik, Karlsruher Institut f\"ur Technologie, 76131 Karlsruhe} 
  \author{T.~Kuhr}\affiliation{Ludwig Maximilians University, 80539 Munich} 
  \author{M.~Heck}\affiliation{Institut f\"ur Experimentelle Kernphysik, Karlsruher Institut f\"ur Technologie, 76131 Karlsruhe} 
  \author{P.~Goldenzweig}\affiliation{Institut f\"ur Experimentelle Kernphysik, Karlsruher Institut f\"ur Technologie, 76131 Karlsruhe} 
  \author{A.~Abdesselam}\affiliation{Department of Physics, Faculty of Science, University of Tabuk, Tabuk 71451} 
  \author{I.~Adachi}\affiliation{High Energy Accelerator Research Organization (KEK), Tsukuba 305-0801}\affiliation{SOKENDAI (The Graduate University for Advanced Studies), Hayama 240-0193} 
  \author{K.~Adamczyk}\affiliation{H. Niewodniczanski Institute of Nuclear Physics, Krakow 31-342} 
  \author{H.~Aihara}\affiliation{Department of Physics, University of Tokyo, Tokyo 113-0033} 
  \author{S.~Al~Said}\affiliation{Department of Physics, Faculty of Science, University of Tabuk, Tabuk 71451}\affiliation{Department of Physics, Faculty of Science, King Abdulaziz University, Jeddah 21589} 
  \author{K.~Arinstein}\affiliation{Budker Institute of Nuclear Physics SB RAS, Novosibirsk 630090}\affiliation{Novosibirsk State University, Novosibirsk 630090} 
% ...
\collaboration{The Belle Collaboration}

\begin{abstract}
We report a measurement of the branching fraction ratios $R(D^{(\ast)})$ of $\Bdecay$ relative to $\Bdecaynorm$ (where $\ell = e$ or $\mu$) using the full Belle data sample of $772 \times 10^6 B\bar{B}$ pairs collected at the $\Upsilon(4S)$ resonance with the Belle detector at the KEKB asymmetric-energy $e^+ e^-$ collider.
The measured values are $R(D)= 0.375 \pm 0.064\mathrm{(stat.)}\pm 0.026\mathrm{(syst.)}$ and $R(D^\ast) = 0.293 \pm 0.038\mathrm{(stat.)}\pm 0.015\mathrm{(syst.)}$.
The analysis uses hadronic reconstruction of the tag-side $B$ meson and purely leptonic $\tau$ decays.
The results are consistent with earlier measurements and do not show a significant deviation from the standard model prediction.
\end{abstract}


\maketitle

{\renewcommand{\thefootnote}{\fnsymbol{footnote}}}
\setcounter{footnote}{0}

\section{Introduction}
Semileptonic $\Bdecaynorm$ decays~\cite{CC}, where $\ell = e$ or $\mu$, have been studied in detail, experimentally~\cite{Agashe:2014kda} and theoretically~\cite{Chay:1990da}, and are used, for example, to extract the standard model (SM) parameter $|V_{cb}|$~\cite{Amhis:2014hma}.
The replacement of the light lepton by the higher-mass $\tau$ 
leads to an increased sensitivity to new physics (NP) effects.
In particular, models with charged Higgs bosons~\cite{Tanaka:1994ay,Itoh:2004ye}, whose couplings are proportional to mass and thus more pronounced for $\tau$ leptons, predict measurable deviations of the branching fraction and kinematic distributions from SM expectations.
The measurement of $\Bdecay$ is challenging because the $\tau$ must be reconstructed from its decay products that include one or more neutrinos.

The first observation of an exclusive semitauonic $B$ decay was reported by the Belle Collaboration in 2007 in the channel $\bar{B}^0 \to D^{\ast+} \tau^- \bar{\nu}_\tau$~\cite{Matyja:2007kt}.
Subsequent measurements by BaBar and Belle~\cite{Aubert:2007dsa,Adachi:2009qg,Bozek:2010xy} reported branching fractions above---yet consistent with---the SM predictions.
In 2012, a significant excess over the SM expectation was reported by BaBar~\cite{Lees:2012xj} that suggested the presence of NP; this called for an independent confirmation.
Interestingly, the two-Higgs doublet model (2HDM) of type II, which might explain a deviation from the SM expectation in a (semi)tauonic $B$ decay~\cite{Tanaka:1994ay}, is incompatible with this result.
A recent LHCb measurement of $\bar{B}^0 \to D^{\ast +} \tau^- \bar{\nu}_\tau$~\cite{Aaij:2015yra} also shows a $2.1\sigma$ deviation from the SM prediction.

Measurements and predictions are usually quoted as branching fraction ratios
\begin{equation}
R(D) = \frac{\mathcal{B}(\BDdecay)}{\mathcal{B}(\BDdecaynorm)}
\end{equation}
and
\begin{equation}
R(D^\ast) = \frac{\mathcal{B}(\BDSdecay)}{\mathcal{B}(\BDSdecaynorm)}
\end{equation}
to reduce experimental systematic uncertainties and theory uncertainties from form factors, where $\mathcal{B}(\Bdecaynorm) = [\mathcal{B}(\bar{B} \to D^{(\ast)} e^- \bar{\nu}_e) + \mathcal{B}(\bar{B} \to D^{(\ast)} \mu^- \bar{\nu}_\mu)] / 2$.
In Ref.~\cite{Lees:2012xj} the calculations in Ref.~\cite{Fajfer:2012vx} are used with updated form factor measurements to obtain the standard model predictions $R(D)_\mathrm{SM} = 0.297 \pm 0.017$ and $R(D^\ast)_\mathrm{SM} = 0.252 \pm 0.003$.
More recent predictions of $R(D)_\mathrm{SM}$ are $0.299 \pm 0.011$~\cite{Lattice:2015rga} and $0.300 \pm 0.008$~\cite{Na:2015kha}.

In this paper, we report new measurements of $R(D)$ and $R(D^\ast)$ with the full Belle $\Upsilon(4S) \to B\bar{B}$ data set of 711~fb$^{-1}$.
The $\tau$ lepton is reconstructed in the leptonic decays $\tau^- \to e^- \bar{\nu}_e \nu_\tau$ and $\tau^- \to \mu^- \bar{\nu}_\mu \nu_\tau$ so that the signal and normalization modes have the same detectable final state particles.
This reduces the systematic uncertainty in $R$ but requires a method to distinguish the modes experimentally.
For this purpose, we exploit the kinematics of $e^+ e^- \to \Upsilon(4S) \to B\bar{B}$ by reconstructing
the accompanying $B$ meson, $B_\mathrm{tag}$, in a hadronic decay mode and extracting the invariant mass squared,
\begin{equation}
\MM = (p_{e^+e^-}-p_{\mathrm{tag}}-p_{D^{(\ast)}}-p_\ell)^2/c^2\ \text{,}
\label{equ:mmiss}
\end{equation}
of all undetected signal-$B$ meson daughters, where $p_\mathrm{e^+e^-}$, $p_{\mathrm{tag}}$, $p_{D^{(\ast)}}$, and $p_\ell$ are the four-momenta of the colliding beam particles, the $B_\mathrm{tag}$ candidate, and the reconstructed signal-$B$ daughters, respectively.

The $\MM$ distribution peaks at (above) zero for the normalization (signal) mode with one neutrino (three neutrinos) in the final state.
The separation power is weaker for backgrounds where multiple final-state particles are not reconstructed.
We improve the rejection of such backgrounds by training a neural network to distinguish them from the signal in the high-$\MM$ region.
Since the low- and high-$\MM$ regions are dominated by different backgrounds, the data sample is split at $\MM=\SI{0.85}{\unitmasssquared}$ and the subsamples are fit simultaneously.
In the low-$\MM$ region, which is dominated by the normalization mode, we fit the $\MM$ distribution; in the high $\MM$ region, where the background with multiple missing particles contributes, we fit the neural-network output distribution.
The analysis procedure is developed and optimized with simulated data before applying it to the experimental data.

\section{Belle Experiment}
This measurement is based on a data sample that
contains $772 \times 10^6 B\overline{B}$ pairs, 
collected  with the Belle detector at the KEKB asymmetric-energy
$e^+e^-$ (3.5 on 8~GeV) collider~\cite{KEKB}
operating at the $\Upsilon(4S)$ resonance.
The Belle detector is a large-solid-angle magnetic
spectrometer that consists of a silicon vertex detector (SVD),
a 50-layer central drift chamber (CDC), an array of
aerogel threshold Cherenkov counters (ACC),
a barrel-like arrangement of time-of-flight
scintillation counters (TOF), and an electromagnetic calorimeter
comprised of CsI(Tl) crystals (ECL) located inside 
a superconducting solenoid coil that provides a 1.5~T
magnetic field.  An iron flux-return located outside of
the coil is instrumented to detect $K_L^0$ mesons and to identify
muons (KLM).  The detector
is described in detail in Ref.~\cite{Belle}.
Two inner-detector configurations were used. A 2.0-cm beampipe
and a three-layer silicon vertex detector was used for the first sample
of $152 \times 10^6 B\bar{B}$ pairs, while a 1.5-cm beampipe, a four-layer
silicon detector, and a small-cell inner drift chamber were used for
the remaining $620 \times 10^6 B\bar{B}$ pairs~\cite{svd2}.

\section{Reconstruction}
We reconstruct $B_\mathrm{tag}$ candidates using the hierarchical hadronic full reconstruction algorithm~\cite{Feindt2011432}, which includes \num{1149} $B$ final states.
The efficiency of the $B_\mathrm{tag}$ reconstruction is 0.3\% for $B^+$ and 0.2\% for $B^0$ mesons~\cite{Feindt2011432}.
Requirements on three observables are applied to enhance the sample's purity: the beam energy-constrained mass $\mbc{}\equiv\sqrt{E_\mathrm{beam}^2-(\mathbf{p}_{\mathrm{tag}}c)^2} / c^2$ must lie between 5.274 and $\SI{5.286}{\unitmass}$, where $E_\mathrm{beam}$ is the colliding-beam energy and $\mathbf{p}_{\mathrm{tag}}$ is the $B_{\textrm{tag}}$ momentum, both measured in the center-of-mass system (CMS);
the absolute value of the energy difference $\Delta E\equiv E_{\mathrm{tag}}-E_\mathrm{beam}$ must be smaller than 50 MeV, where $E_{\mathrm{tag}}$ is the $B_{\textrm{tag}}$ CMS energy; and the full-reconstruction neural-network quality estimator for $B_{\textrm{tag}}$
(which incorporates modified Fox-Wolfram moments~\cite{KSFW} to suppress $e^+ e^- \to q\bar{q}$ continuum events)
must exceed a channel-dependent threshold that
preserves $\approx 85\%$ of the $\Bdecay$ events. 

In each event with a selected $B_{\textrm{tag}}$ candidate, we search for the signature $D^{(\ast)}\ell$, with $\ell=e$ or $\mu$, among the remaining tracks and calorimeter clusters.
The four disjoint data samples are denoted $D^+\ell^-$, $D^0\ell^-$, $D^{\ast+}\ell^-$, and $D^{\ast0}\ell^-$.
We reconstruct $D^+$ mesons in the decays to $K^-\pi^+\pi^+$, $K_S^0\pi^+$, $K_S^0\pi^+\pi^0$, and $K_S^0\pi^+\pi^+\pi^-$; $D^0$ mesons to $K^-\pi^+$, $K^-\pi^+\pi^+\pi^-$, $K^-\pi^+\pi^0$, $K_S^0\pi^0$, and $K_S^0\pi^+\pi^-$; $D^{\ast +}$ mesons to $D^0\pi^+$ and $D^+\pi^0$; and $D^{\ast 0}$ mesons to $D^0\pi^0$ and $D^0\gamma$.

Charged-particle candidates are selected from tracks that originate from within \SI{4.0} (\SI{2.0}) cm along (perpendicular to) the beam direction of the interaction point (IP).
Selections on the particle-identification likelihood ratio of the electron (muon) vs. the hadron hypothesis for the candidate lepton track retain \SI{95}{\percent} (\SI{92}{\percent}) of signal events.
We veto a $D^{(\ast)}$ candidate if a charged daughter is lepton-like, with a signal efficiency of \SI{97}{\percent}.
$K_S^0$ candidates are reconstructed from pairs of oppositely charged tracks, treated as pions, and must satisfy standard quality requirements~\cite{Sumisawa:2005fz}. 

Clusters in the ECL with an energy of at least \SI{50}{\unitlowenergy} and no matching track are identified as photons.
Candidate $\pi^0$'s are reconstructed from pairs of photons. 
For end-cap photons used in a $\pi^0$ candidate, the energy must be greater than 80~MeV.
The momentum of $\pi^0$ candidates not originating from a $D^\ast$ decay must exceed \SI{200}{\unitlowmomentum} in the signal-$B$ rest frame. 
The absolute value of the difference $S_{\gamma\gamma}$ between the invariant mass of the $\pi^0$ candidate and the nominal $\pi^0$ mass, normalized to its uncertainty, must be below \num{3.0}. 

We select $D^{+/0}$ meson candidates with a CMS momentum below \SI{3.0}{\unitmomentum}.
For both $D$ and $D^*$ candidates, the candidate $D$ mass and $D^*-D$ mass difference must be within 1.5 standard deviations of the nominal $D$ mass and $D^*-D$ mass difference, respectively.
The resolution is asymmetric and is taken from simulated data.

The missing mass squared, $\MM$, must lie between \num{-0.2} and $\SI{8.0}{\unitmasssquared}$.
The momentum transfer $q^2 \equiv (p_{B} - p_{D^{(\ast)}})^2$ on the signal side is required to be greater than \SI{4.0}{\unitmomentumsquared}, which suppresses the otherwise overwhelming contribution from semileptonic $B$ meson decays to light leptons.
Events with a remaining $\pi^0$ candidate are rejected if the energy of either daughter photon exceeds \num{50}/\num{100}/\SI{150}{\unitlowenergy} in the barrel/forward/backward region.
The overall charge of the event must be zero, with no additional charged tracks allowed.

If there are several $B_{\textrm{tag}}$ candidates, the one with the most signal like neural-network quality estimator is selected. 
Then, on average, we have \num{1.23} signal or normalization candidates per event, with most ambiguities arising from $D^\ast$ meson decays to a $D$ meson and a neutral pion or photon in $\BDSdecaynorm$ decays.
In a multicandidate event, we select one at random.

\section{Simulation}
We use samples of simulated (MC) events to study backgrounds, to optimize the selection criteria, and to determine the probability density function (PDF) shapes of the fit components.
The decay chains in all simulated data are generated with the EvtGen~\cite{Lange:2001sv} package; the GEANT3~\cite{brun1987geant3} framework is used to simulate the detector response.
A luminosity-weighted run-dependent sample of \num{e+7} events for each of the four signatures is generated for the signal mode $\Bdecay$ using the decay model described in Ref.~\cite{Tanaka:2012nw}.
To investigate possible new physics effects, we produce a sample of simulated $\Bdecay$ signal events for the scenario of a two-Higgs-doublet model of type II with $\tan{\beta}/m_{H^+}=\SI{0.5}{\clight\squared/\giga\electronvolt}$~\cite{Tanaka:2012nw}.
A sample that corresponds to 5 times the amount of recorded data and contains $B\bar{B}$ events with $B$ mesons decaying generically via $b \to c$ transitions as well as $q\bar{q}$ events with $q \in \{u,d,s,c\}$ is used for the background.

Several corrections are applied to the background MC sample to improve its agreement with measured data.
We first reweight the MC events to account for the imperfect estimate of the proportions of correctly reconstructed $B_\mathrm{tag}$ candidates and to better estimate 
the yields of background processes with good tags.
(The reweighting cancels to first order in the efficiency ratio used to extract $R(D^{(\ast)})$.) 
The weights are given by the ratios of yields in simulation and data, determined from fits to the distributions of $\mbc{}$ and $\MM$ for events with a $B_\mathrm{tag}$ and a semileptonic decay on the signal side~\cite{Sibidanov:2013rkk}.
The correction factors are in the range 0.35 to 1.1, with an overall factor of approximately 0.75.
To extract correction factors for the number of incorrectly reconstructed $B_{\textrm{tag}}$ candidates, we compare yields of simulated and reconstructed data in a sideband of $\mbc{}$, requiring $\SI{5.23}{\unitmass}<\mbc{}<\SI{5.25}{\unitmass}$.
This is done separately for the four signal modes, and we exclude events with fake $D^{(\ast)}$ mesons or fake leptons on the signal side as these are corrected by other measures.
The ratios of the yields, whose values are between 0.99 and 1.14, are then applied as weights.

Second, we apply a correction for the signal-lepton candidates to account for differing misidentification rates in simulated and recorded data.
Correction factors for the lepton candidate are provided in eight (eleven) bins in polar angle (momentum).
(Lepton-identification efficiencies are compatible, within uncertainties, between simulated and recorded data.)

Third, we reweight the events to account for $D^{(\ast)}$ yield differences in MC and data.
While the yield of candidates with a fake $D$ meson will be estimated from sidebands and therefore does not need to be corrected in simulated data, differences in correctly reconstructed $D$ yields can affect the determination of $R$.
We determine the yield ratios of simulated and reconstructed data by fitting the invariant mass (mass difference) distributions of the $D$ ($D^{\ast}$) mesons in a wider window than used for the nominal selection and apply the ratios as weights.
This is done individually for each $D^{(\ast)}$ meson reconstruction channel and yields correction factors between 0.75 and 1.09.
Background MC events with $D_s^- \to \ell^- \bar{\nu}_\ell$ decays are reweighted to adopt the latest branching fraction measurements~\cite{Agashe:2014kda}.

Fourth, semileptonic decays of $B$ mesons to higher excitations of $D$ mesons, hereinafter labelled $D^{\ast\ast}$, comprise one of the most challenging backgrounds.
Our background MC sample contains semileptonic---including semitauonic---$B$ decays to $D_2^\ast$, $D_0^\ast$, $D_1$, $D_1^\prime$, and the radial excitations $D(2S)$ and $D^\ast(2S)$, each in the charged and neutral variety.
The decays are generated initially according to the ISGW model~\cite{ISGW} and reweighted to reproduce the distributions in $\QSQ$ and $\PLEP$ (the lepton momentum in the signal-$B$ frame) of the LLSW model~\cite{LLSW}.
Parameter uncertainties in this model are treated as systematic uncertainties.
We consider $D^{\ast\ast}$ decays to a $D^{(\ast)}$ and one or two pions, a $\rho$, or an $\eta$ meson, with branching ratio assumptions based on quantum-number, phase-space, and isospin arguments.
Similar weights are applied to $\Bdecaynorm$ events in the background MC
according to the most recent measurements of the form factors $\rho^2=1.207 \pm 0.015 \pm 0.021$, $R_1=1.403 \pm 0.033$, and $R_2=0.854 \pm 0.020$ for $\BDSdecaynorm$ and $\rho^2=1.186 \pm 0.036 \pm 0.041$ for $\BDdecaynorm$~\cite{Caprini,Amhis:2014hma}.

\section{Sample Composition}
We identify the following components in the data samples:
\begin{description}
\item[lepton normalization]
This originates from $\Bdecaynorm$ decays and has both visible (i.e., non-neutrino) daughters of the $B$ meson correctly reconstructed
with a distinctive $\MM$ distribution that peaks around zero.
Its yield is a free parameter of the fit.
\item[lepton cross-feed]
This arises from the misclassification of a
$\BDSdecaynorm$ decay into the $D\ell^-$ sample with same $D$-meson charge due to the loss of a low-energetic $\pi^0$ or $\gamma$ daughter of the $D^\ast$.
The broad $\MM$ distribution peaks at positive values up to roughly
$\SI{1.0}{\unitmasssquared}$.
Its yield is allowed to float in the fit.
\item[tau signal]
This component, arising from $\Bdecay$ decays, has a correctly reconstructed $D^{(\ast)}$ daughter and a correctly identified $\tau$-decay daughter lepton.
With three final-state neutrinos, its broad $\MM$ distribution is
most prominent in regions above $\SI{1.0}{\unitmasssquared}$.
The yield $Y^{D\ell^-}_{\tau\,\mathrm{signal}}$ of the \component{tau signal} in each $D\ell^-$ sample is determined by the branching-fraction ratio $R(D)$, which is a free parameter in the fit, the corresponding \component{lepton normalization} yield $Y^{D\ell^-}_{\ell\,\mathrm{norm}}$, and the efficiency ratio $f^D$ for the \component{lepton normalization} and \component{tau signal} components:
\begin{equation}
Y^{D^{+,0}\ell^-}_{\tau\,\mathrm{signal}} = R(D)
Y^{D^{+,0}\ell^-}_{\ell\,\mathrm{norm}}/(2f^{D^{+,0}}).
\end{equation}
The factor of 2 accounts for the inclusion of both electrons and muons in the \component{lepton normalization} component.
The efficiency ratios, which include the $\tau^- \to \ell^- \nu_\tau \bar{\nu}_\ell$ branching fractions~\cite{Agashe:2014kda}, are determined from simulation to be $f^{D^+} = 1.69\pm0.09$ and $f^{D^0} = 1.91\pm0.06$, where the uncertainties are statistical.
In a similar way, the \component{tau signal} yield in the $D^\ast\ell^-$ samples is given by the floating fit parameter $R(D^\ast)$ and the corresponding \component{lepton normalization} yield.
However, to encompass larger yields and thus obtain smaller statistical uncertainties, the \component{cross-feeds} are added to the \component{tau signal} and \component{lepton normalization} with the concomitant use of an effective efficiency ratio $f^{D^{\ast}}_\mathrm{eff}$, defined by
\begin{equation}
\frac{1}{f^{D^{\ast +,0}}_\mathrm{eff}} = \frac{1-x_\mathrm{CF}}{f^{D^{\ast +,0}}}  + \frac{x_\mathrm{CF}}{f^{D^{\ast +,0}}_\mathrm{CF}} \, ,
\end{equation}
where $x_\mathrm{CF}$ is the fraction of \component{lepton cross-feed} events relative to the sum of \component{lepton normalization} and \component{lepton cross-feed} yields, determined from simulation, and $f^{D^{\ast}}$ ($f^{D^{\ast}}_\mathrm{CF}$) is the efficiency ratio for the \component{lepton normalization} (\component{lepton cross-feed}) and \component{tau signal} (\component{tau cross-feed}) components.
The values of the effective efficiency ratios are $f^{D^{\ast +}}_\mathrm{eff} = 3.11\pm0.13$ and $f^{D^{\ast 0}}_\mathrm{eff} = 3.63\pm0.09$.
\item[tau cross-feed]
This component is the analogue to the \component{lepton cross-feed} but originating from $\BDSdecay$ decays.
Its yield and shape in $\MM$ are quite similar to those of the \component{tau signal} component.
It appears only in the $D\ell^-$ samples and its yield is constrained by the \component{$\tau$ signal} yield $Y^{D^{\ast}\ell}_{\tau\,\mathrm{signal}}$ in the respective $D^\ast\ell^-$ samples of same charge, assuming a $\pi^0$ or $\gamma$ from the $D^\ast$ decay is not reconstructed.
The constraining factor is taken from the appropriate \component{lepton normalization} and \component{lepton cross-feed} yields and is calibrated by a factor $g$ that represents the cross-feed ratio for light-lepton and $\tau$ modes; MC gives $g^+ = 0.83\pm0.08$ for the $D^+\ell^-$ sample and $g^0 = 0.69\pm0.04$ for the $D^0\ell^-$ sample.
The \component{tau cross-feed} yield $Y^{D\ell}_{\tau\,\mathrm{CF}}$ is given by
\begin{equation}
Y^{D^{+,0}\ell^-}_{\tau\,\mathrm{CF}} = Y^{D^{\ast+,0}\ell^-}_{\tau\,\mathrm{signal}} 
\frac{Y^{D^{+,0}\ell^-}_{\ell\,\mathrm{CF}}}{Y^{D^{\ast+,0}\ell^-}_{\ell\,\mathrm{norm}}} \frac{1}{g^{+,0}}.
\end{equation}
\item[wrong-charge lepton cross-feed]
This component is similar to \component{lepton cross-feed} but arises from the loss of the charged pion in
$D^{\ast +} \to D^0\pi^+$.
To preserve the overall neutral charge of the event, the lost pion is absorbed into the now-misreconstructed
$B_\mathrm{tag}$ meson.
(Since $D^{\ast 0}$ mesons do not decay to charged pions, this component appears only in the $D^0\ell^-$ sample.)
Its $\MM$ distribution resembles that of \component{lepton cross-feed}.
Its smaller yield is constrained in the fit relative to the \component{lepton normalization} yield in the $D^{\ast+}$ sample with a factor $f_\mathrm{wc}=0.107\pm0.004$, taken from simulation.
\item[fake \boldmath$D^{(\ast)}$]
This component is dominated by random combinations of final-state particles that form a fake $D$ or $D^\ast$ meson.
This can happen by either missing particles in the event or misassigning particles to the wrong $B$ meson.
This background occurs in all samples and, in the $D^\ast\ell^-$ samples, includes combinations of a correctly reconstructed $D$ meson and an incorrect $D^\ast$ primary daughter.
The $\MM$ distribution is very broad and extends to the highest values.

The \component{fake $D^{(\ast)}$} yield is estimated separately for each $D^{(\ast)}$ decay mode.
A sideband region is defined in the distribution of the invariant mass $M_D$ (the $D^\ast$--$D$ mass difference $\Delta M_{D^\ast D}$) by excluding twice the signal-region width on both sides of the nominal mass (mass difference) and a $\pm 60$~MeV$/c^2$ window around the $D^\ast$ peak position for the $D^+\to K_S^0\pi^+\pi^0$ channel.
Multiplying the sideband yield in the real data by the yield ratio in MC of the \component{fake $D^{(\ast)}$} component and sideband provides an estimate for the \component{fake $D^{(\ast)}$} yield in each $D^{(\ast)}$ decay mode; these are summed to obtain the total yield in each of the four data samples.
\item[\boldmath$D^{\ast \ast}$ background]
This component contains candidates that originate from $\bar{B}\to D^{\ast\ast}\ell^-\bar{\nu}_\ell(\nu_\tau\bar{\nu}_\tau)$ decays.
The higher-excitation $D$ states decay typically to a $D$ or $D^\ast$ meson plus one pion (although more pions are possible)
so the final state here has a properly identified lepton, a properly reconstructed $D^{(\ast)}$ meson, and (at least) one pion that might be lost or absorbed into $B_{\textrm{tag}}$.
If the pion is missed, this process mimics the \component{tau signal} and exhibits a similar $\MM$ distribution.
The yields of this background and the \component{tau signal} are comparable.
In contrast to the other background components, it is not possible to constrain the yield from MC since the properties of $\bar{B}\to D^{\ast\ast}\ell^-\bar{\nu}_\ell$ and $D^{\ast\ast}$ decays are not known reliably.
Thus, its yield is a free parameter in the fit.
\item[fake lepton] 
This component contains events with a misidentified lepton candidate; the track
is usually a kaon or pion from the tag side or from $\bar{B}\to D K$ or $D\pi$ decays.
This component also includes $\Bdecay$ events with a misidentified pion from a hadronic $\tau$ decay.
Since lepton misidentification is far less probable than $D^{(\ast)}$ misreconstruction, a $D^{(\ast)}\ell^-$ event in which both are misidentified 
is classified in the \component{fake $D^{(\ast)}$} component.
The \component{fake lepton} background is a broad structure in $\MM$ that appears in all four data samples; the fixed and relatively low yield is estimated accurately from MC.
\item[\boldmath$D_s$ decay]
This component arises from the decay chain $\bar{B}\to D^{(\ast)}D_s^-$ with $D_s^- \to \ell^- \bar{\nu}_\ell (\nu_\tau \bar{\nu}_\tau)$ and so has a final state that mimics the \component{tau signal}.
The decay $D_s^- \to \ell^- \bar{\nu}_\ell$ is helicity suppressed and only the tauonic $D_s^-$ decays provide a non-negligible contribution.
Its $\MM$ distribution resembles that of the \component{tau signal}; its low yield in MC is confirmed by experiment, with the most precise determination provided by Belle~\cite{Zupanc:2013byn}.
Consequently, this component's yield is fixed in the fit to the MC value.
\item[rest]
This component encompasses all background candidates that are not captured by the other listed components.
It contains candidates with well-identified final state particles that do not originate from one of the previously covered sources and may be random combinations of tag- and signal-side particles.
Its yield is quite low in all four samples and is fixed in the fit to the MC value.
\end{description}
Table~\ref{tab:fit:yield_overview} itemizes each component in the fit for each signature.
The yields of the fixed components are listed in Table~\ref{tab:fit:expectations:expectations}.

\begin{table}[htb]
\centering
\caption{Fit components in each data sample. For the yield source, ``fit'' indicates a free parameter in the fit; ``constrained'' reflects a dependence on other parameters; ``MC'' denotes a fixed yield taken from simulation; and ``SB'' identifies a fixed yield derived from the corresponding sideband. The constraints are described in the text.}
\label{tab:fit:yield_overview}
\begin{ruledtabular}
\begin{tabular}{lccccc} 
	{Component}        & {$D^+\ell^-$}        & {$D^0\ell^-$}              & {$D^{\ast+}\ell^-$}  & {$D^{\ast0}\ell^-$}        & {Yield source}     \\ \hline
	$\ell$ normalization  & \checkmark           & \checkmark                 & \checkmark           & \checkmark                 & Fit                       \\
	$\ell$ CF          & \checkmark           & \checkmark                 & -                    & -                          & Fit                       \\
	$\tau$ signal      & \checkmark           & \checkmark                 & \checkmark           & \checkmark                 & Fit                       \\
	$\tau$ CF          & \checkmark           & \checkmark                 & -                    & -                          & Constrained               \\
	Wrong charge $\ell$ CF    & -                    & \checkmark                 & -                    & -                          & Constrained               \\
	Fake $D$           & \checkmark           & \checkmark                 & -                    & -                          & $M_D$ SB        \\
	Fake $D^\ast$      & -                    & -                          & \checkmark           & \checkmark                 & $\Delta M_{D^\ast D}$ SB \\
	$D^{\ast\ast}$ background & \checkmark           & \checkmark                 & \checkmark           & \checkmark                 & Fit                       \\
	Fake $\ell$        & \checkmark           & \checkmark                 & \checkmark           & \checkmark                 & MC                        \\
	$D_s$ decay        & \checkmark           & \checkmark                 & \checkmark           & \checkmark                 & MC                        \\
	Rest               & \checkmark           & \checkmark                 & \checkmark           & \checkmark                 & MC 
\end{tabular}
\end{ruledtabular}
\end{table}

\begin{table}[htb]
\centering
\caption{Yields for the fixed components in the four data samples.
}
\label{tab:fit:expectations:expectations}
\begin{ruledtabular}
\begin{tabular}{lrrrr}
	                   & $D^+\ell^-$ & $D^0\ell^-$ & $D^{\ast+}\ell^-$ & $D^{\ast0}\ell^-$ \\ \hline
	Fake $D^{(\ast)}$  &         350 &        1330 &               180 &              2220 \\
	Fake $\ell$        &        20.9 &          69 &              13.7 &              12.9 \\
	$D_s$ decay        &        22.0 &         112 &              21.0 &              20.7 \\
	Rest               &        23.6 &          77 &               4.3 &               4.2 
\end{tabular}
\end{ruledtabular}
\end{table}

\section{Fit Procedure}
As explained above, the low-$\MM$ region is dominated by the \component{lepton normalization} and has essentially no sensitivity to the \component{tau signal}; in contrast, the high-$\MM$ region, where the \component{tau signal} is concentrated, exhibits little discrimination power in $\MM$ between the \component{tau signal} and the other backgrounds---in particular, the \component{$D^{\ast\ast}$ background}.
Therefore, we fit simultaneously the $\MM$ distribution below $\SI{0.85}{\unitmasssquared}$ to constrain the \component{lepton normalization} and \component{lepton cross-feed} yields and a neural-network output $o_{\rm NB}$ above $\SI{0.85}{\unitmasssquared}$ to constrain the yields of the other components.  (In fact, \emph{all} components are fit in both regions.)
The partition at $\MM=\SI{0.85}{\unitmasssquared}$ minimizes the expected uncertainty on $R(D)$ and $R(D^\ast)$.

The aforementioned neural network is trained for each of the four data samples with simulated events to distinguish the \component{tau signal} from the backgrounds in the high-$\MM$ region: mainly \component{$D^{\ast \ast}$ background} but also the \component{wrong-charge cross-feed}, \component{fake lepton}, \component{$D_s$ decay}, and \component{rest} components.
The neural network incorporates $\MM$ and several other observables that provide the desired signal-to-background separation.
The most powerful observable is $\ecl$, the unassociated energy in the ECL that aggregates all clusters that are not associated with reconstructed particles (including bremsstrahlung).
A nonzero $\ecl$ value indicates a missing physical process in the event, such as a decay mode with a $\pi^0$ in which only a single daughter photon is reconstructed.
Two additional network inputs are $q^2$ and $\PLEP$; their additional discriminating power is limited by their strong correlation with $\MM$.
Other input variables, which provide marginally more discrimination, are the number of unassigned $\pi^0$ candidates with $|S_{\gamma\gamma}| < \num{5.0}$; the cosine of the angle between the momentum and vertex displacement of the $D^{(\ast)}$ meson; and the decay-channel identifiers of the $B$ and $D^{(\ast)}$ mesons.

For use in the fit, the neural-network output $\NBO$ is transformed into
\begin{equation}
  \NBTR \equiv \log \frac{\NBO - o_{\text{min}}}{o_{\text{max}} - \NBO}\ ,
\end{equation}
where the parameters $o_{\text{min}}$ and $o_{\text{max}}$ are the minimum and maximum network output values, respectively, in the elected data sample.
The $\NBTR$ distributions have smoother shapes and can be described well with bifurcated Gaussian functions, which makes their parameterizations more robust.

For each fit component within a selected data sample, two PDFs are determined: in $\MM$ for $\MM<\SI{0.85}{\unitmasssquared}$ and in $\NBTR$ for $\MM>\SI{0.85}{\unitmasssquared}$.
The PDFs of $\MM$ are represented by smoothed histograms and
are constructed by applying a smoothing algorithm~\cite{splft} to the respective MC distributions.
Each bifurcated-Gaussian PDF in $\NBTR$ is parameterized by the mean, left width and right width, 
which are determined by an unbinned maximum likelihood fit to the MC distribution.
In the fit, each component has a total yield, defined in Table~\ref{tab:fit:yield_overview}, with partial yields in the lower- and upper-$\MM$ regions that are fixed MC-determined fractions of the total yield.

We maximize the extended likelihood function
\begin{equation}
\mathcal{L}=\prod\limits_i\left[\mathcal{Q}(N_i,K_i)\prod\limits_{k_i=1}^{K_i} \mathcal{P}_i(\mathbf{x}_{k_i})\right] \ ,
\end{equation}
where $i \in \lbrace D^+\ell^-,\, D^0\ell^-,\, D^{\ast+}\ell^-,\, D^{\ast0}\ell^-\rbrace$ is the data-sample index, $\mathcal{Q}(N_i,K_i)$ is the Poisson probability to observe $K_i$ events for an expectation value of $N_i=\sum\limits_{j}Y_{i,j}$ events (with $Y_{i,j}$ being the yield of component $j$ in data sample $i$), and the vector $\mathbf{x}_{k_i}$ holds the values for $\MM$ and $\NBTR$ of candidate $k_i$.
The PDF $\mathcal{P}_i$ of data sample $i$ is given by
\begin{eqnarray}
  \nonumber
  \mathcal{P}_i(\MM,\NBTR) &=& \frac{1}{N_i} \cdot 
  \sum\limits_{j}Y_{i,j} \big{[} f_{i,j,\mathrm{low}}  \mathcal{P}_{i,j,\mathrm{low}}(\MM) + \\*
  &&(1-f_{i,j,\mathrm{low}}) \mathcal{P}_{i,j,\mathrm{high}}(\NBTR)
       \big{]} \ .
\end{eqnarray}
The index $j$ runs over the components and $f_{i,j,\mathrm{low}}$ is the fraction of events of the component $j$ that are in the lower $\MM$ range. The one-dimensional probability density function $\mathcal{P}_{i,j,\mathrm{low}}$ ($\mathcal{P}_{i,j,\mathrm{high}}$) represents the $\MM$ ($\NBTR$) distribution in the low- (high-)$\MM$ region.

The simultaneous fit over all four data samples has twelve free parameters: the \component{lepton normalization} yield per sample, the \component{lepton cross-feed} yield per $D\ell^-$ sample, the \component{$D^{\ast\ast}$ background} yield per sample, and the branching-fraction ratios $R(D)$ and $R(D^\ast)$.
Here, we assume isospin symmetry and use the same $R(D)$ and $R(D^\ast)$ parameters for the $\bar{B}^0$  and $B^-$ samples.

\section{Cross-Checks}
The implementation of the fit procedure is tested by applying the same procedure to multiple subsets of the available simulated data.
The fit accuracies are evaluated using sets of \num{500} pseudoexperiments and show no significant bias in any measured quantity.
These are used also to test the influence on the fit result of the value of $\MM=\SI{0.85}{\unitmasssquared}$ that is used to partition the samples:
variation of this value reduces the precision of the fit result but does not introduce any bias.

Further tests address the compatibility of the simulated and recorded data.
To test resolution modelling, we use a sample of events with $q^2<3.5$~GeV$^2/c^2$, dominated by $\Bdecaynorm$ decays.
As the \component{$D^{\ast\ast}$ background} is one of the most important components---with a large potential for flaws in its modeling---we evaluate its distributions in more depth by reconstructing a data sample with enriched $\bar{B}\to D^{\ast\ast}\ell^-\bar{\nu}_\ell$ content by requiring a signal-like event but with an additional $\pi^0$.
The background-enriched data samples are fit individually in four dimensions separately: $\MM$, $\MMR$, $\ecl$, and $\PLEP$, where $\MMR$ is the missing mass of the candidate, calculated without the additional $\pi^0$.
The shapes of the components are extracted from simulated data.
In each of the four $D^{(\ast)}\ell^-\pi^0$ samples, consistent yields are obtained from the fits to all four variables, indicating that the simulation describes faithfully the distribution in all tested dimensions.

\section{Results}
\begin{figure*}[htbp]
	\includegraphics[width=\columnwidth]{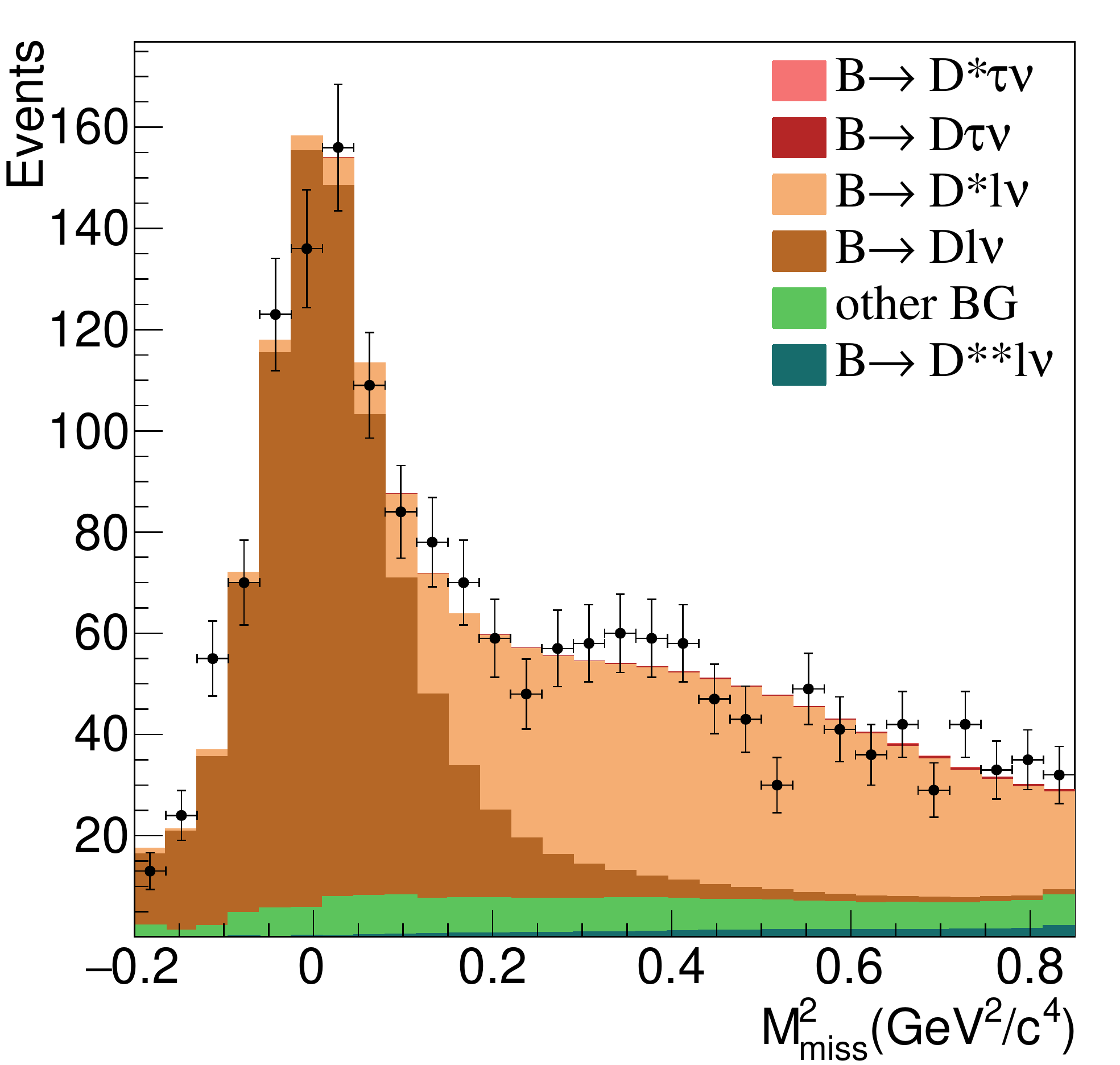}
	\includegraphics[width=\columnwidth]{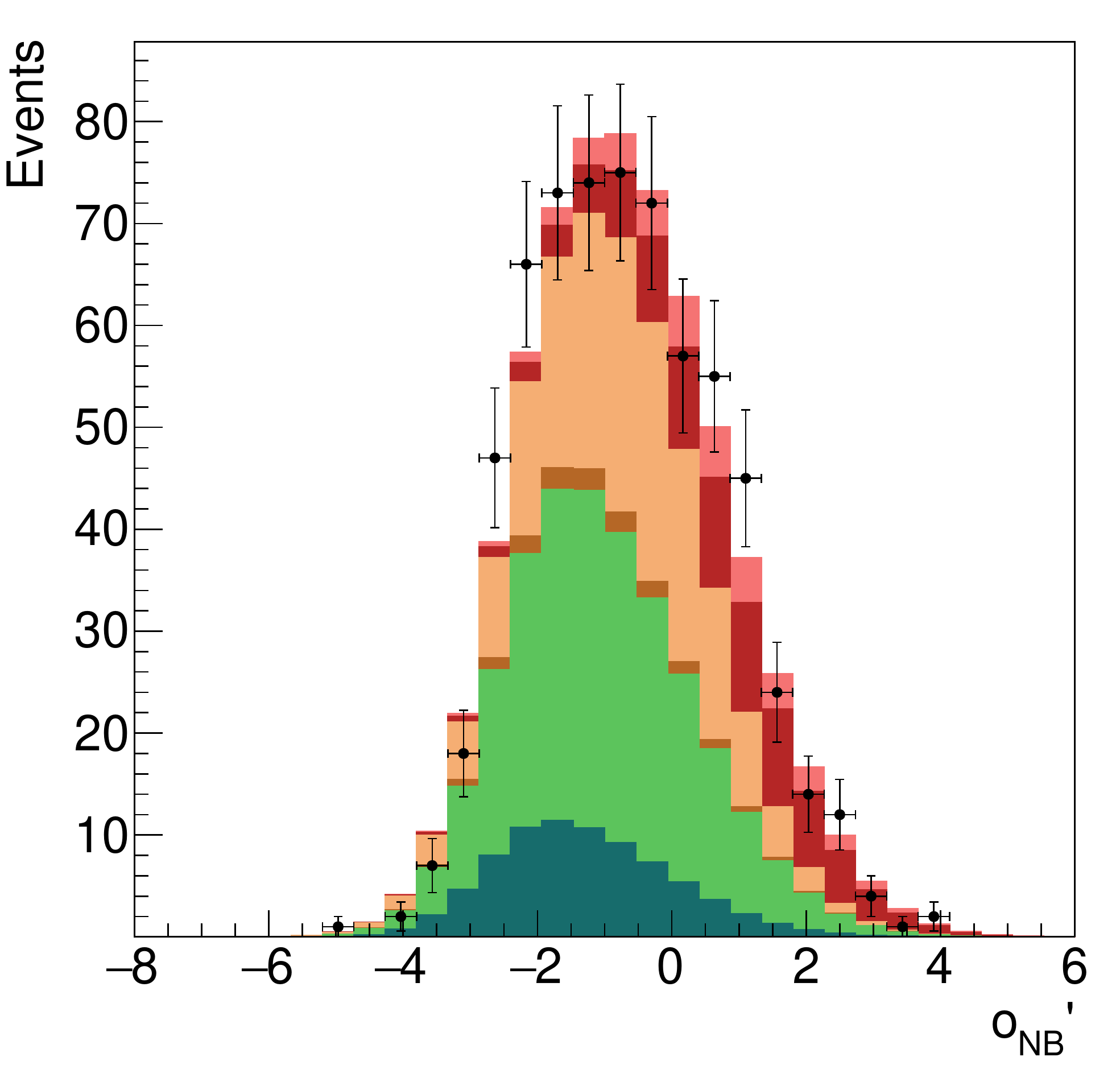}
	\includegraphics[width=\columnwidth]{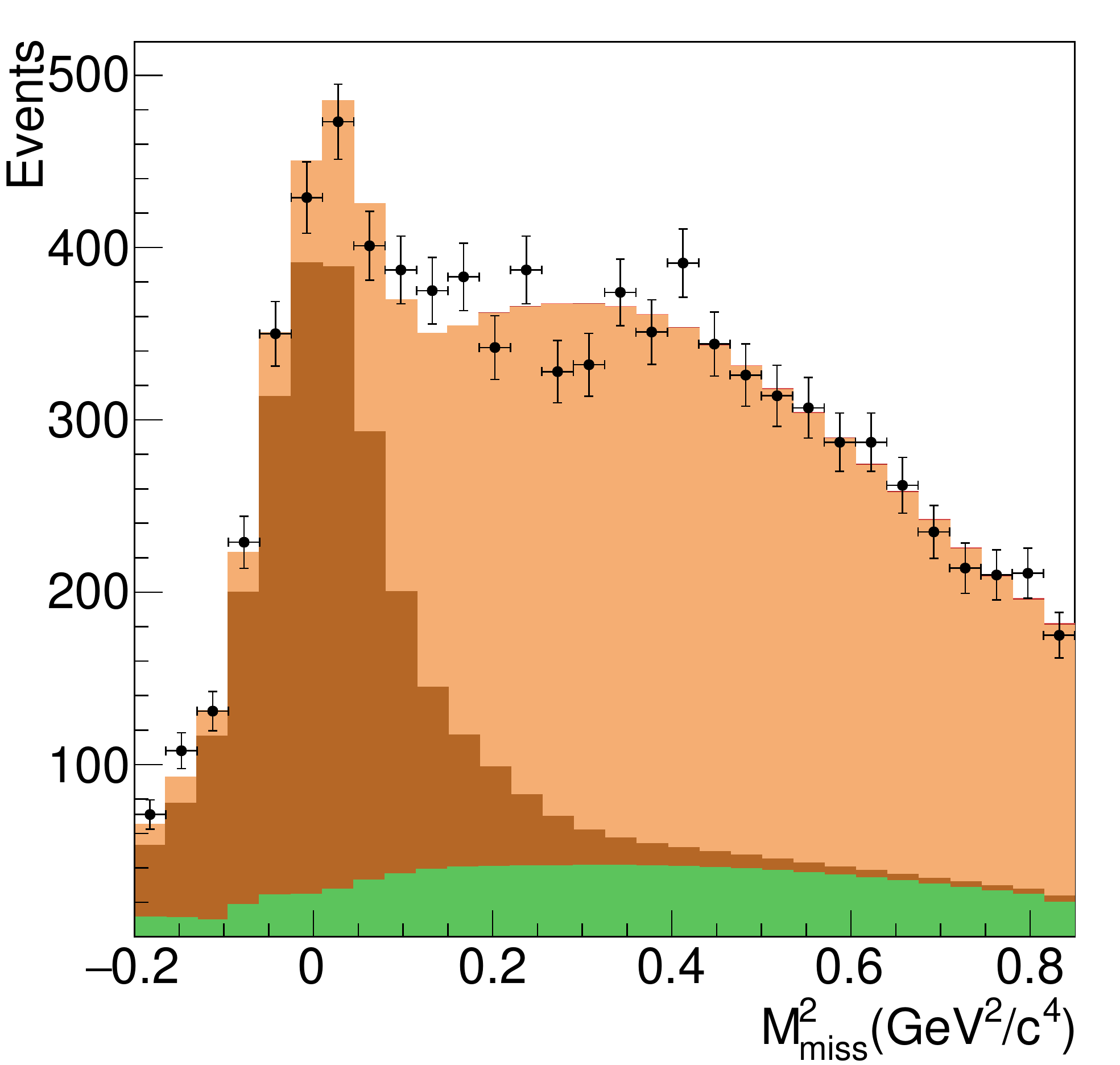}
	\includegraphics[width=\columnwidth]{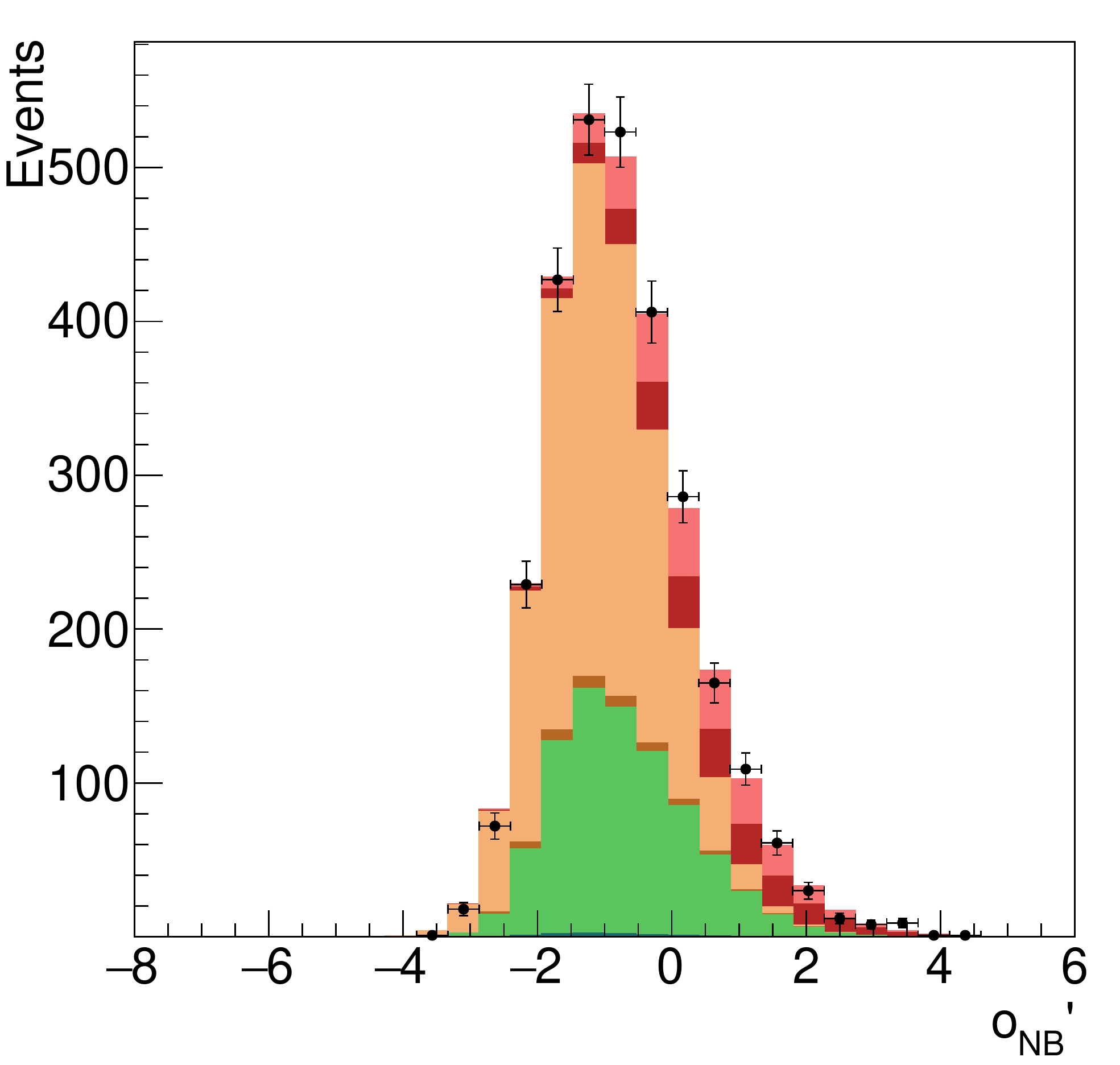}
\caption{Fit projections and data points with statistical uncertainties in the $D^+\ell^-$ (top) and $D^0\ell^-$ (bottom) data samples.
Left: $\MM$ distribution for $\MM<\SI{0.85}{\unitmasssquared}$; right: $\NBTR$ distribution for $\MM>\SI{0.85}{\unitmasssquared}$. }
\label{fig:results:Dmodes}
\end{figure*}

\begin{figure*}[htbp]
	\includegraphics[width=\columnwidth]{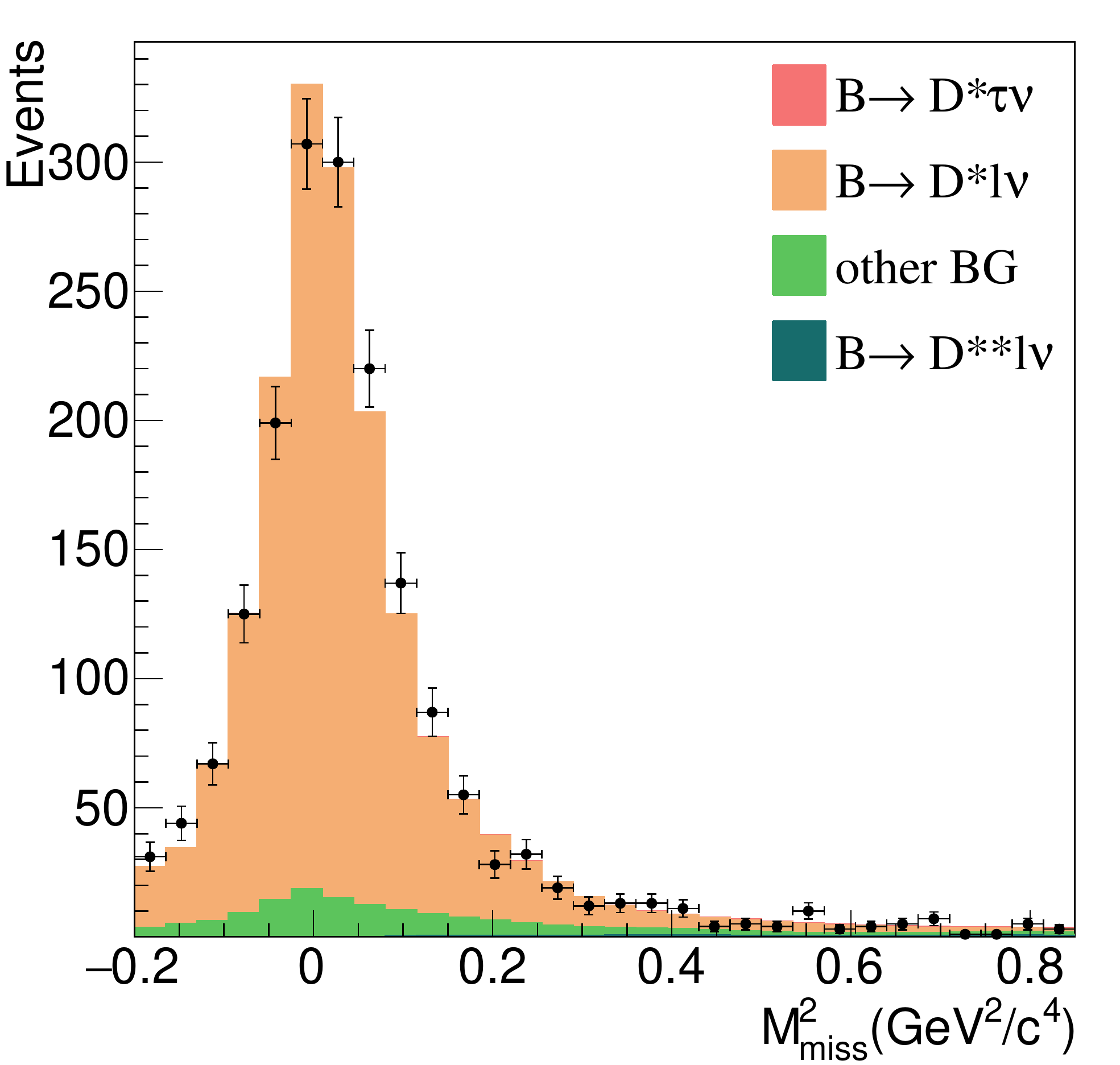}
	\includegraphics[width=\columnwidth]{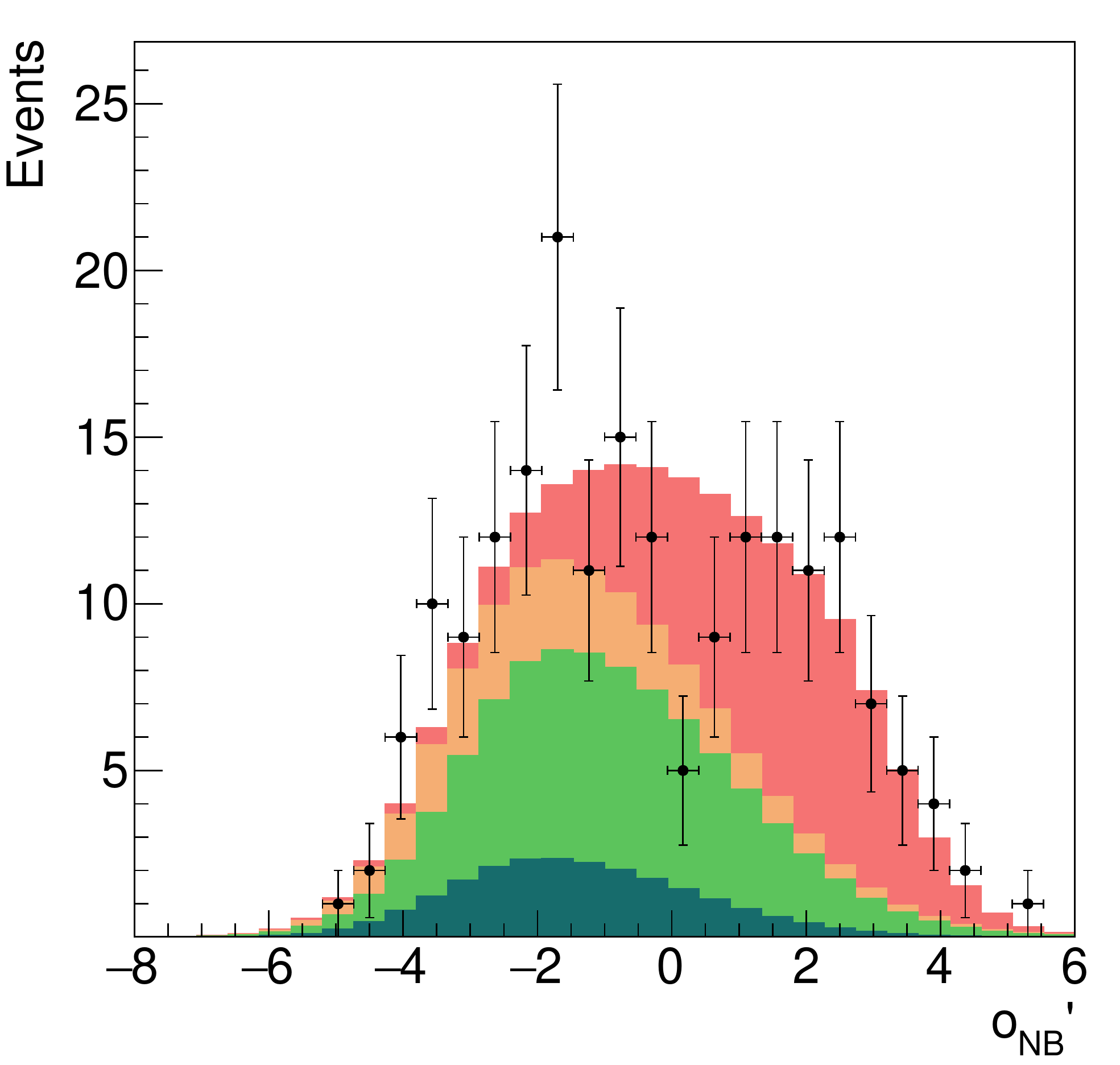}
	\includegraphics[width=\columnwidth]{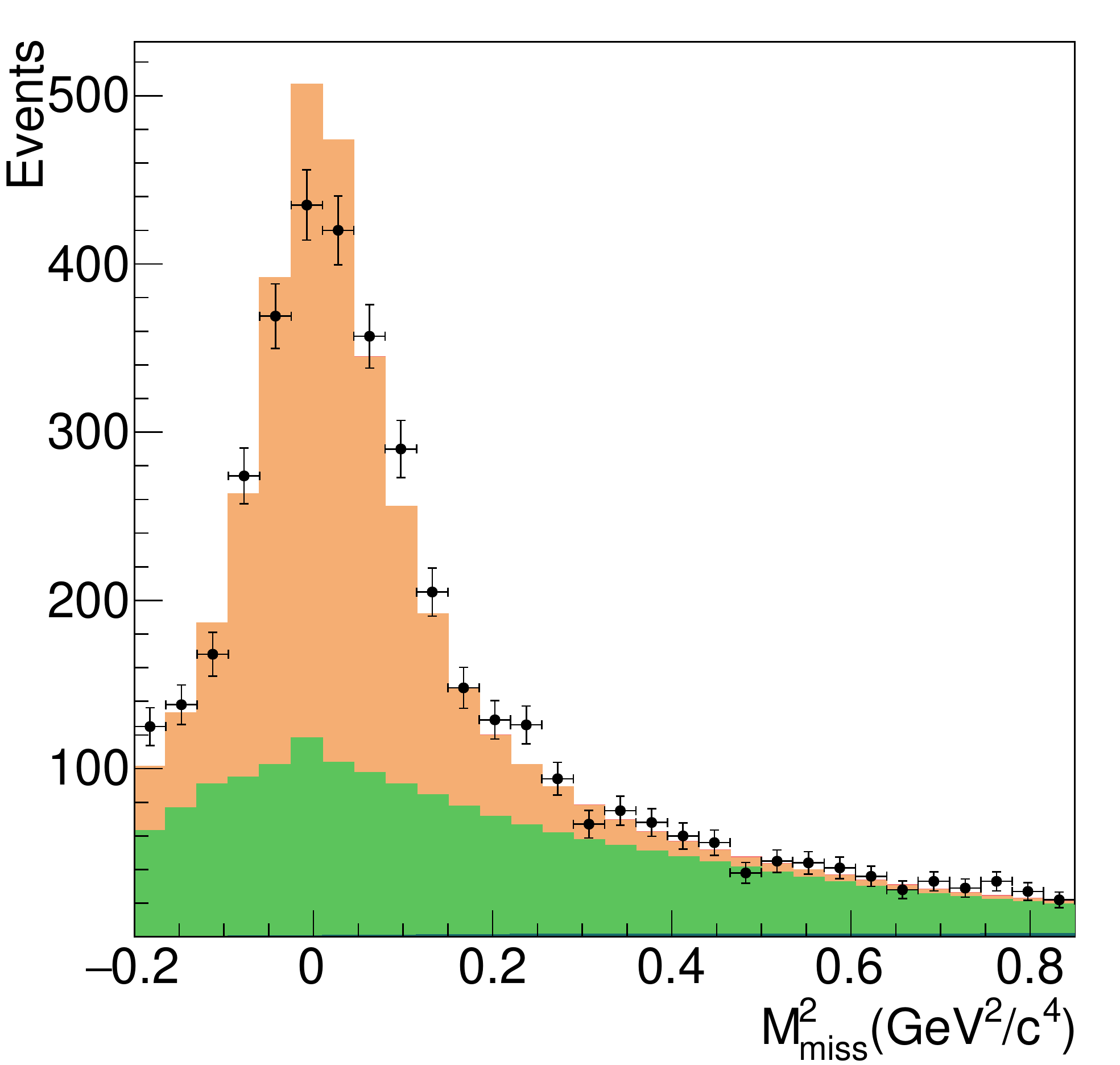}
	\includegraphics[width=\columnwidth]{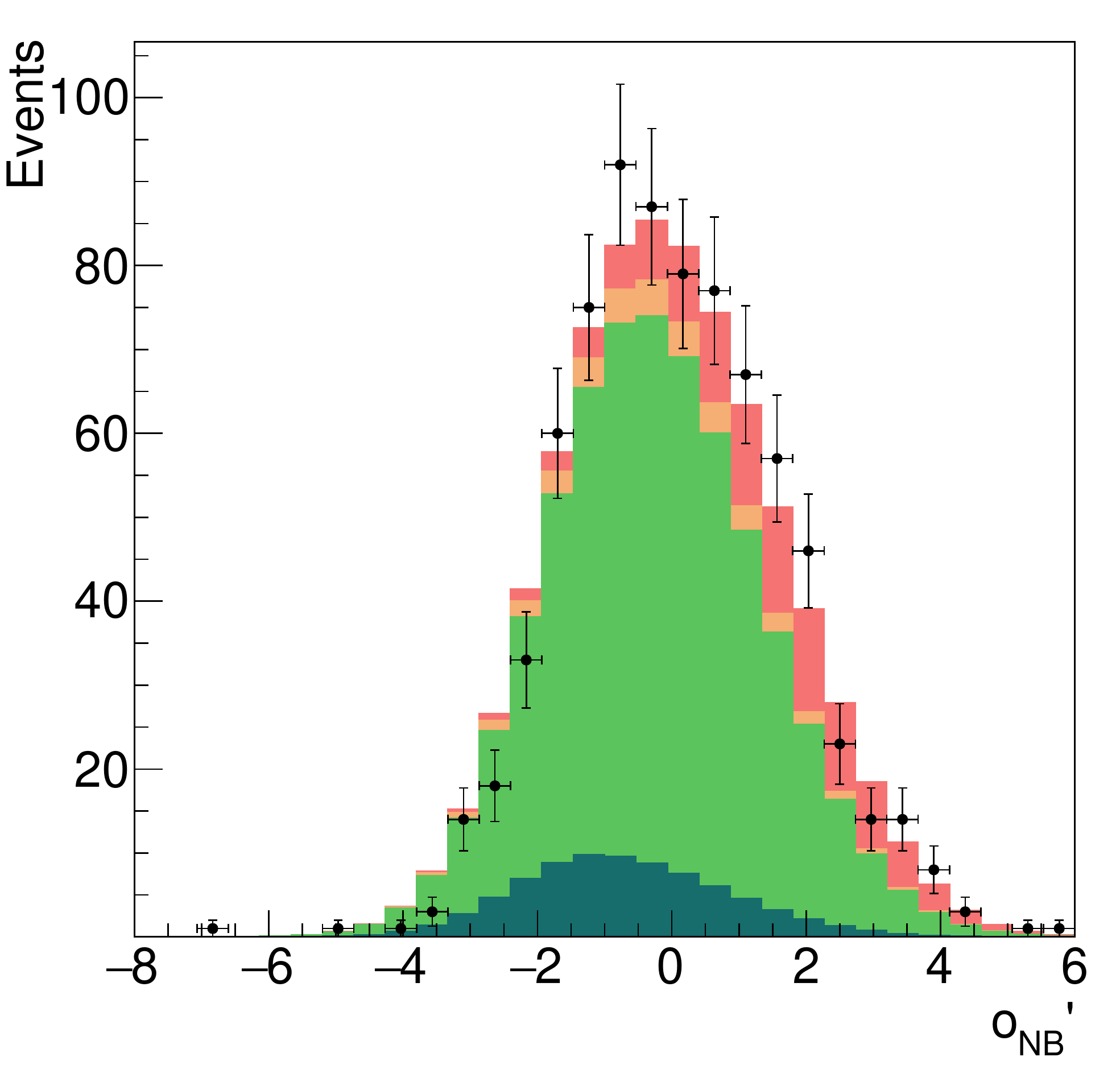}
\caption{Fit projections and data points with statistical uncertainties in the $D^{\ast +}\ell^-$ (top) and $D^{\ast 0}\ell^-$ (bottom) data samples.
Left: $\MM$ distribution for $\MM<\SI{0.85}{\unitmasssquared}$; right: $\NBTR$ distribution for $\MM>\SI{0.85}{\unitmasssquared}$. }
\label{fig:results:Dstmodes}
\end{figure*}

\begin{figure*}[htbp]
	\includegraphics[width=\columnwidth]{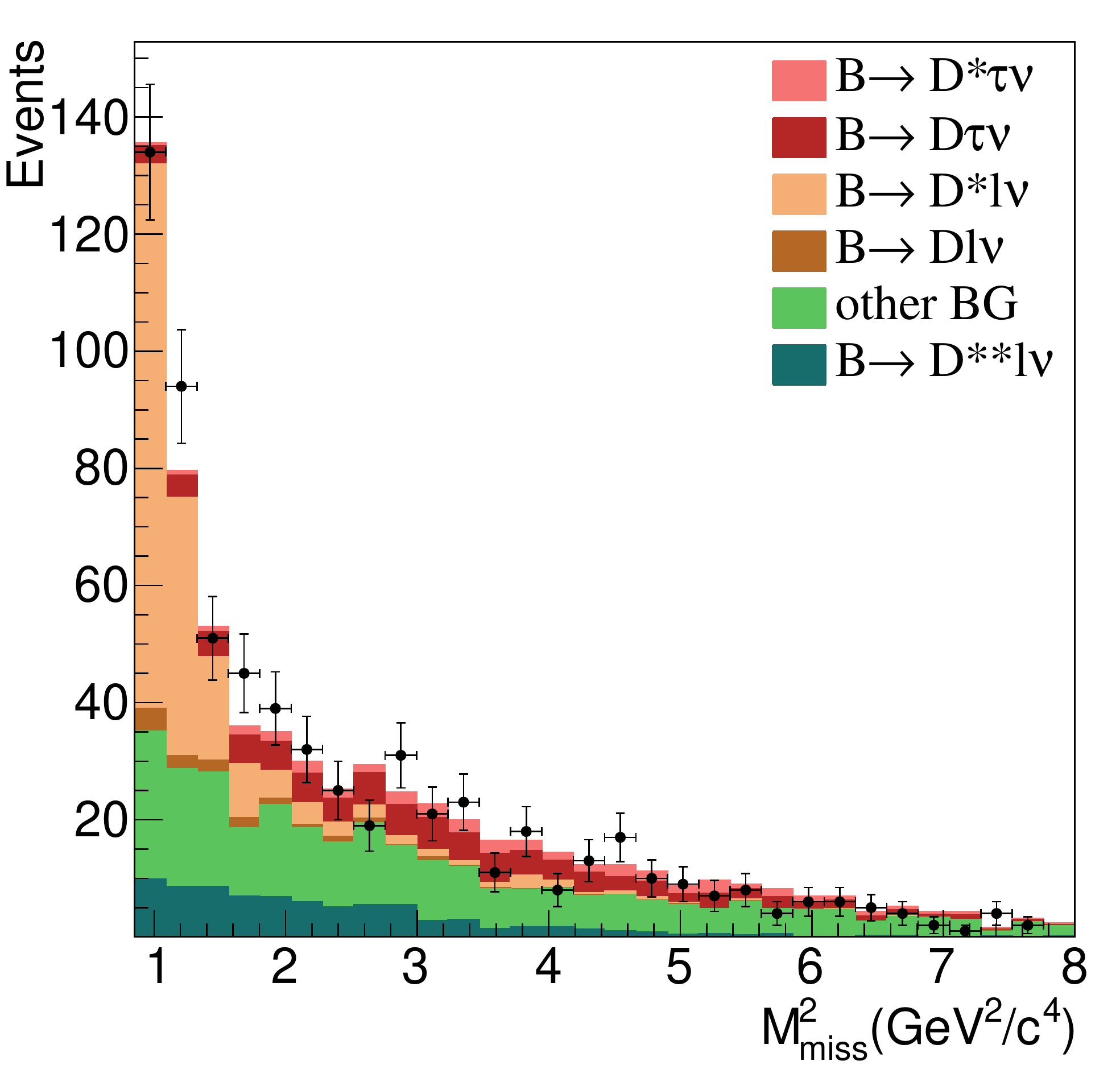}
	\includegraphics[width=\columnwidth]{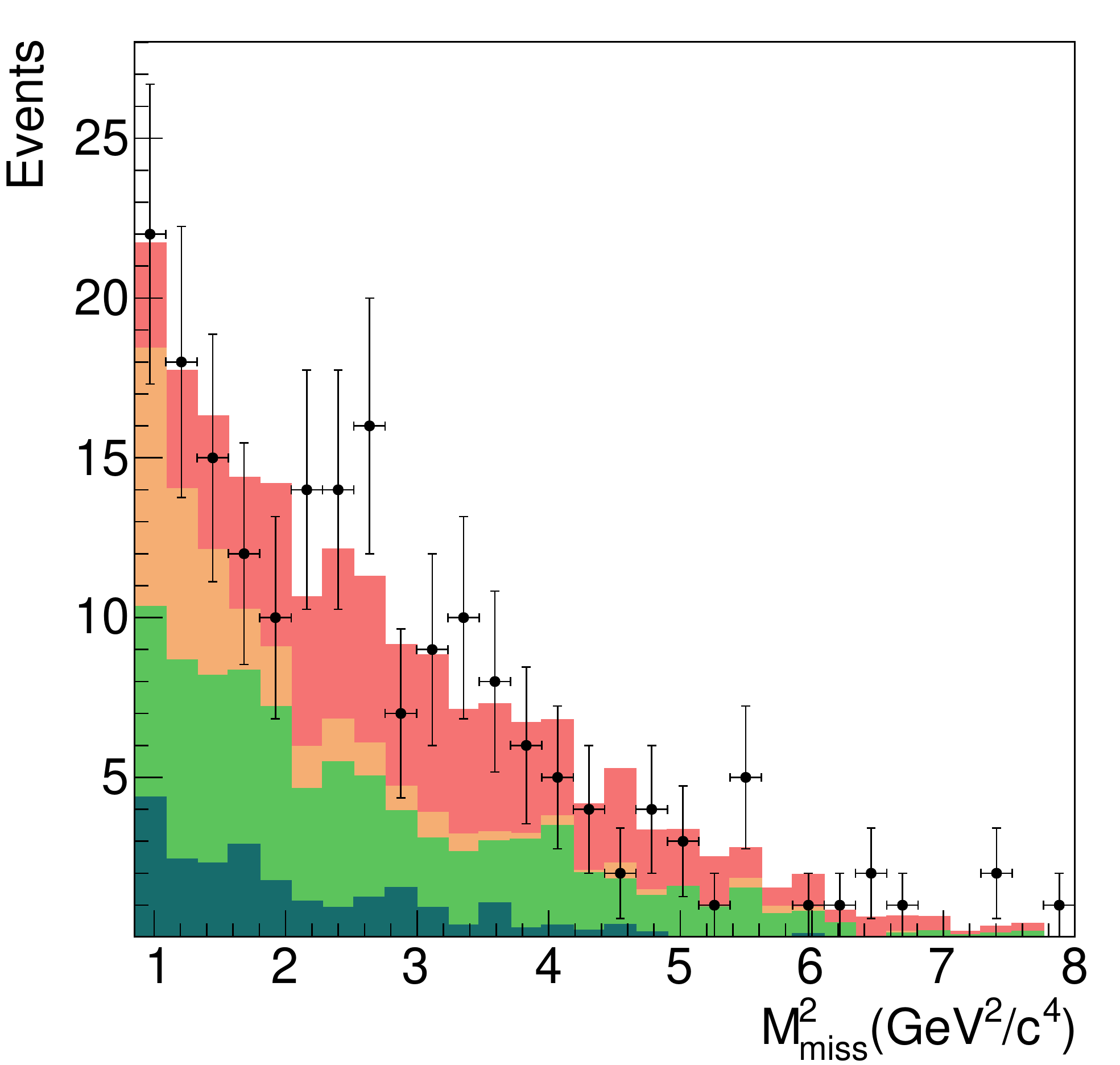}
	\includegraphics[width=\columnwidth]{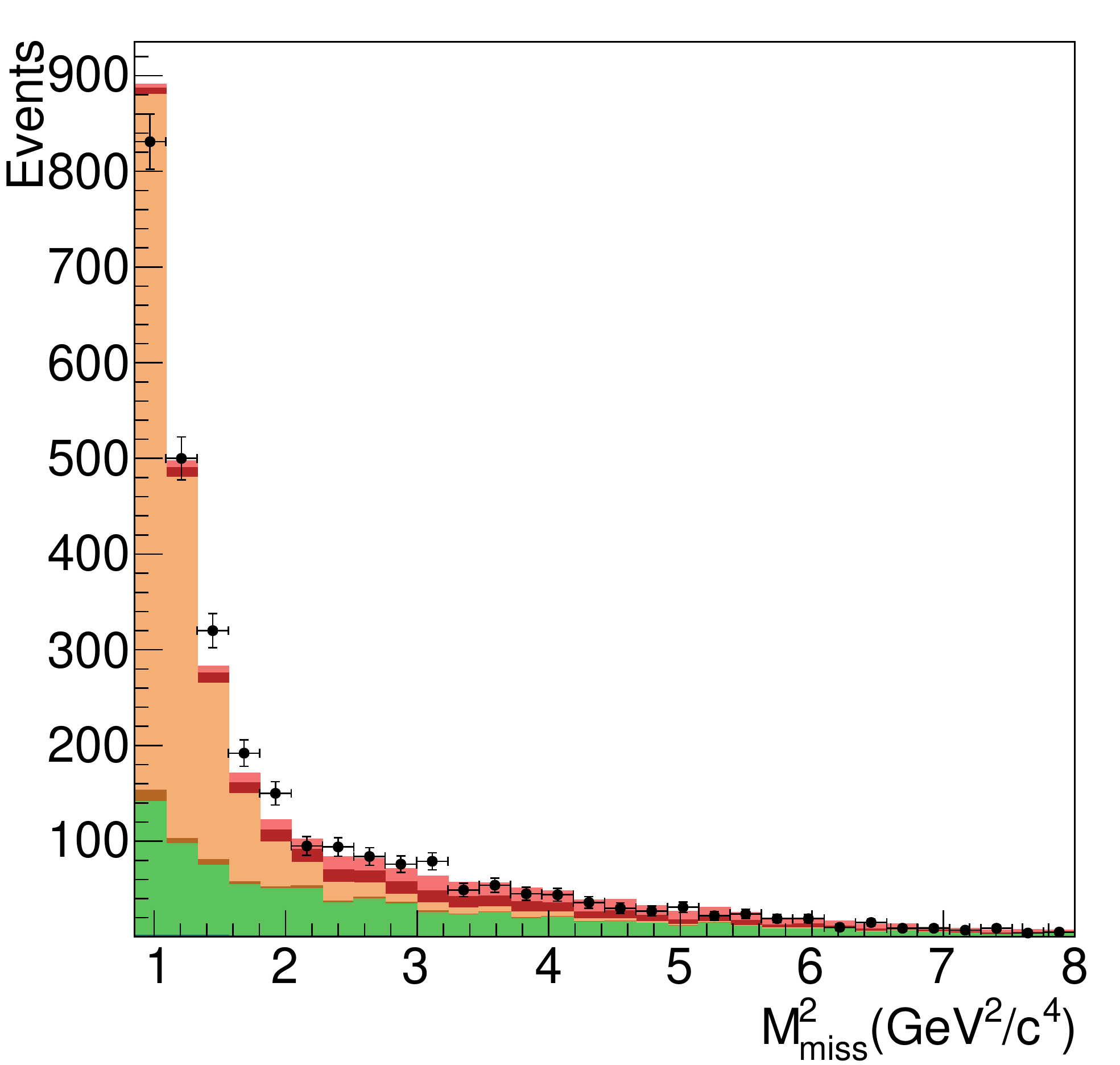}
	\includegraphics[width=\columnwidth]{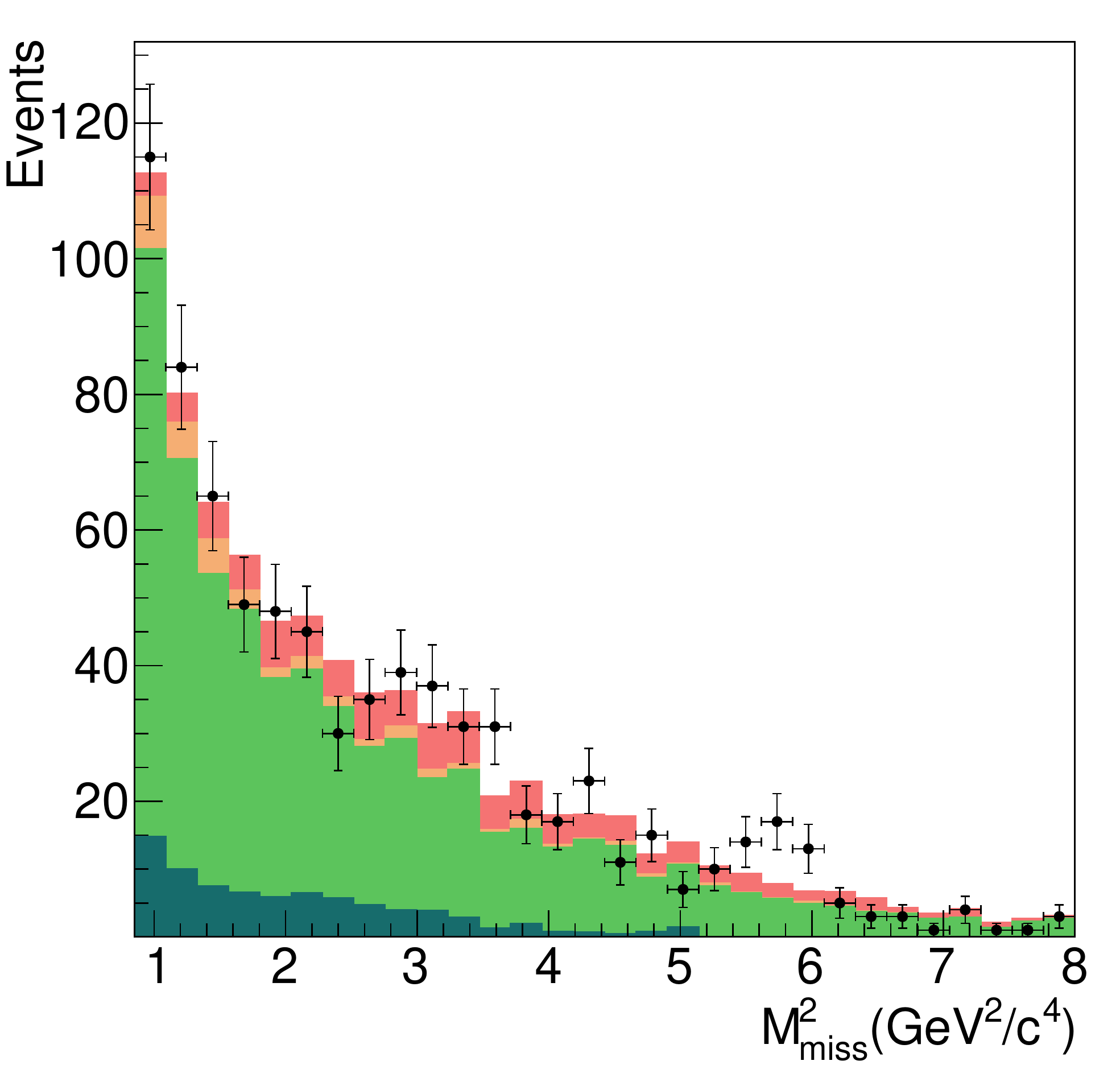}
\caption{Projections of the fit results and data points with statistical uncertainties for the high $\MM$ region.
Top left: $D^{+}\ell^-$; top right: $D^{\ast +}\ell^-$; bottom left: $D^{0}\ell^-$; bottom right: $D^{\ast 0}\ell^-$.}
\label{fig:results:control_mm}
\end{figure*}

\begin{figure*}[htbp]
	\includegraphics[width=\columnwidth]{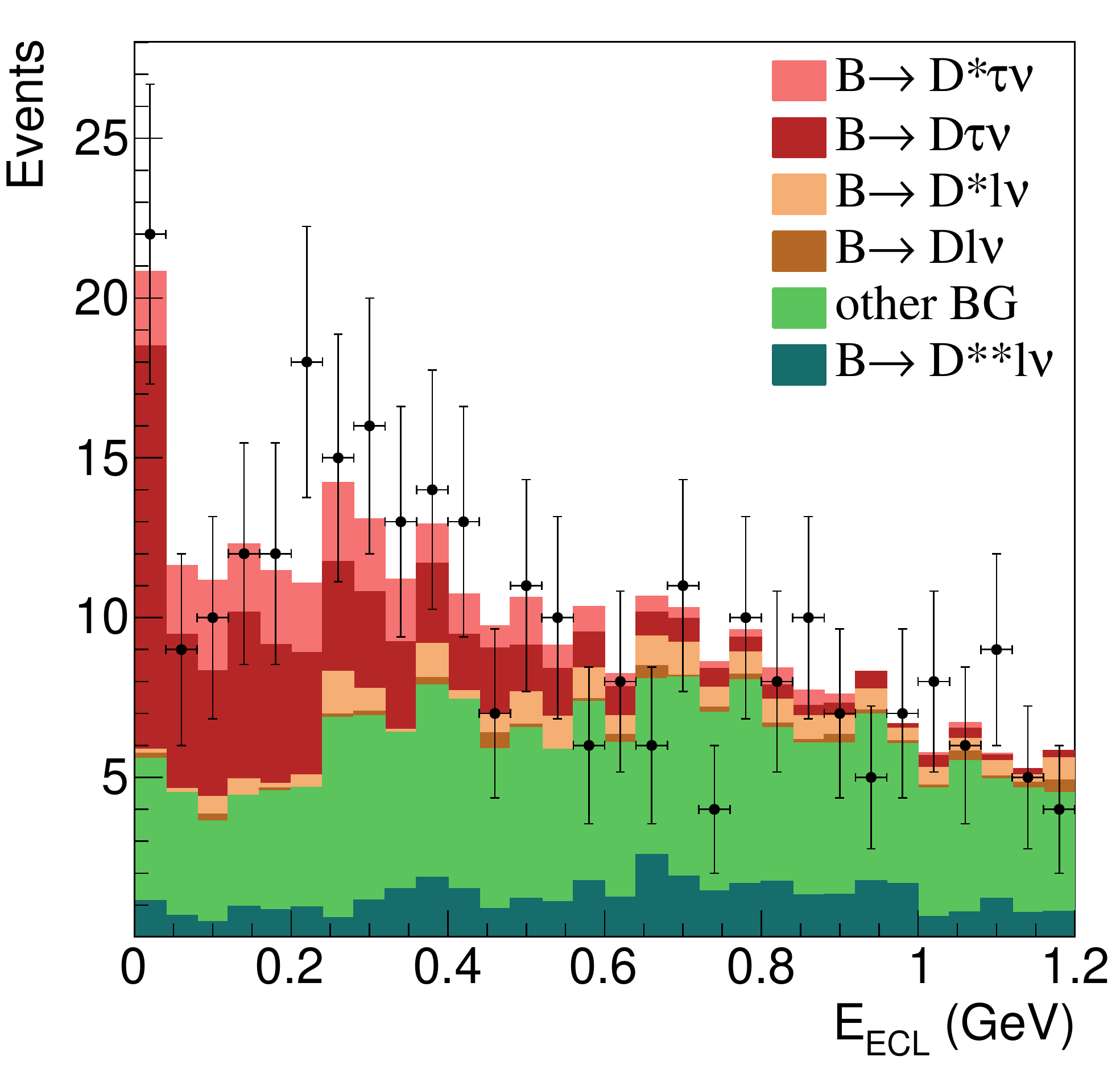}
	\includegraphics[width=\columnwidth]{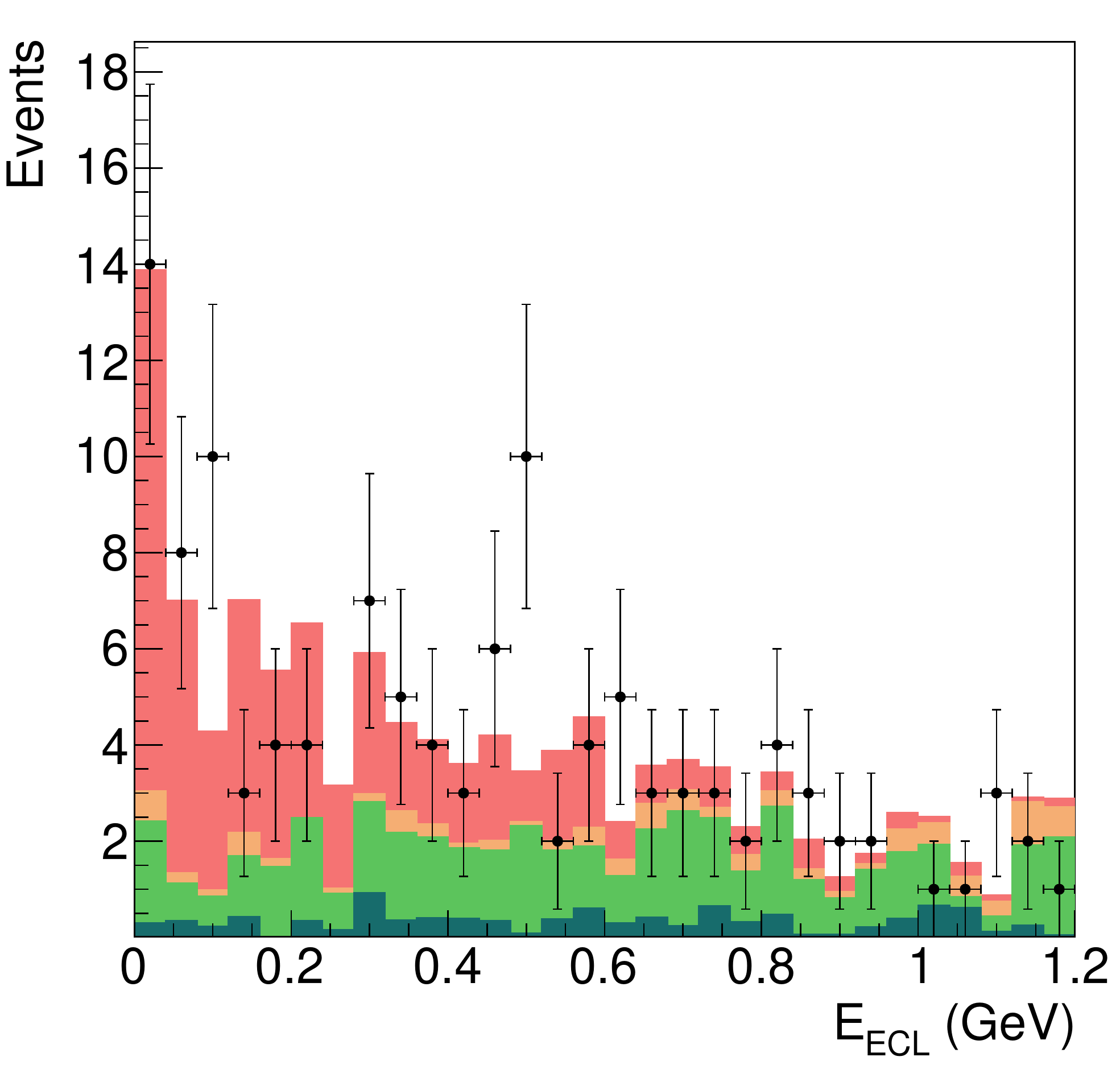}
	\includegraphics[width=\columnwidth]{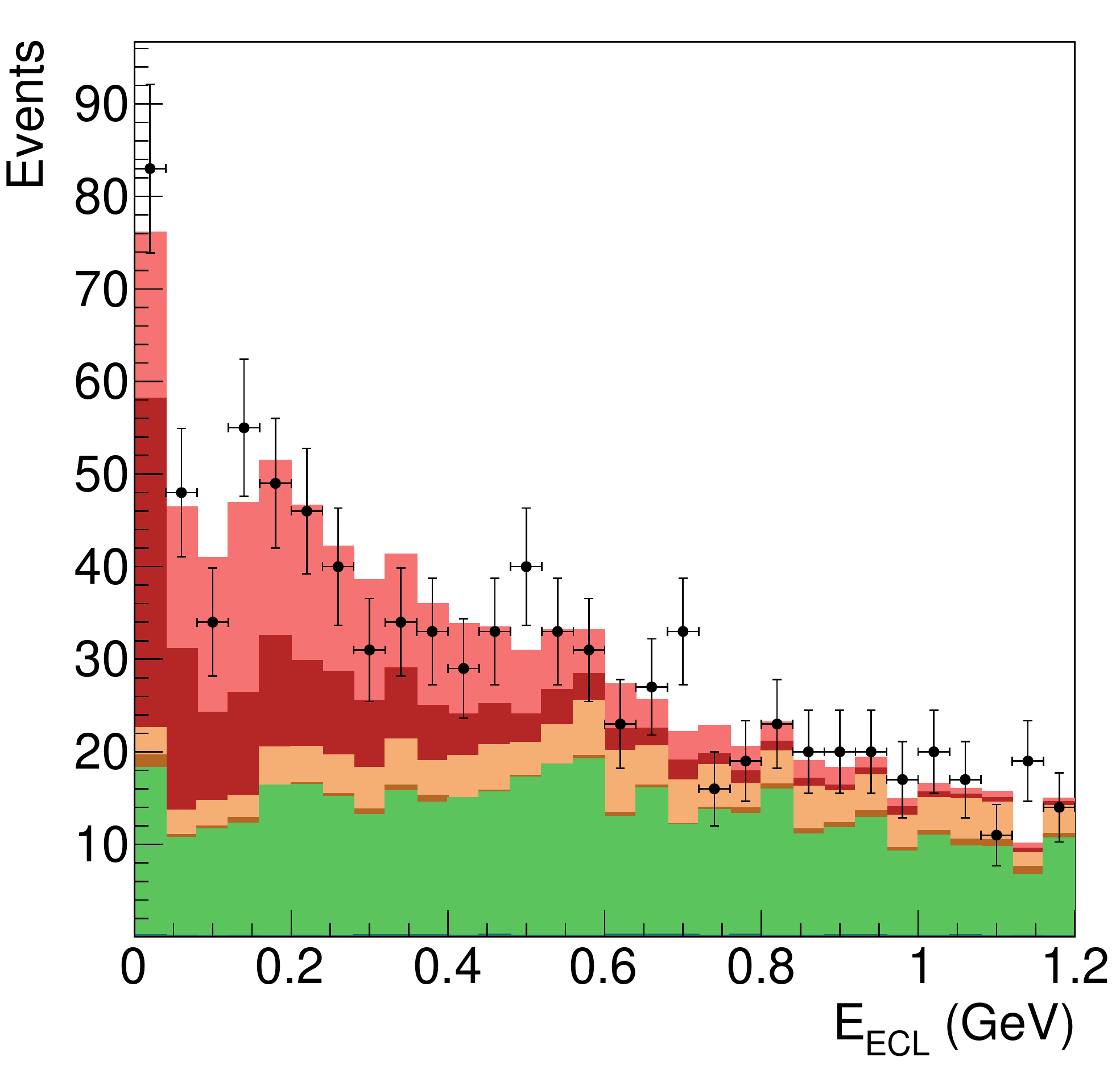}
	\includegraphics[width=\columnwidth]{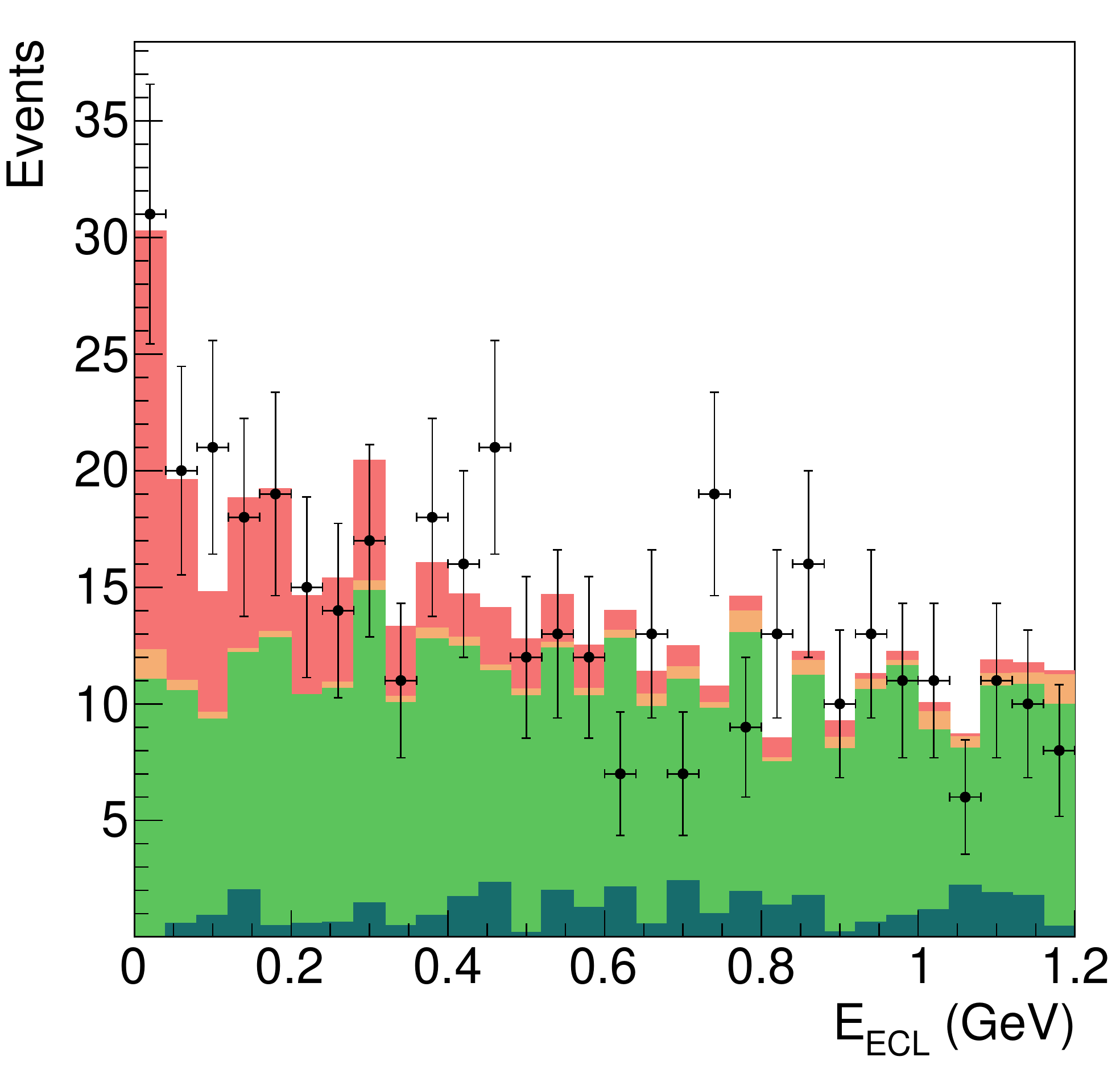}
\caption{Projections of the fit results and data points with statistical uncertainties in a signal-enhanced region of $\MM>\SI{2.0}{\unitmasssquared}$ in the $\ecl$ dimension.
Top left: $D^{+}\ell^-$; top right: $D^{\ast +}\ell^-$; bottom left: $D^{0}\ell^-$; bottom right: $D^{\ast 0}\ell^-$.}
\label{fig:results:control_eecl}
\end{figure*}

\begin{figure*}[htbp]
	\includegraphics[width=\columnwidth]{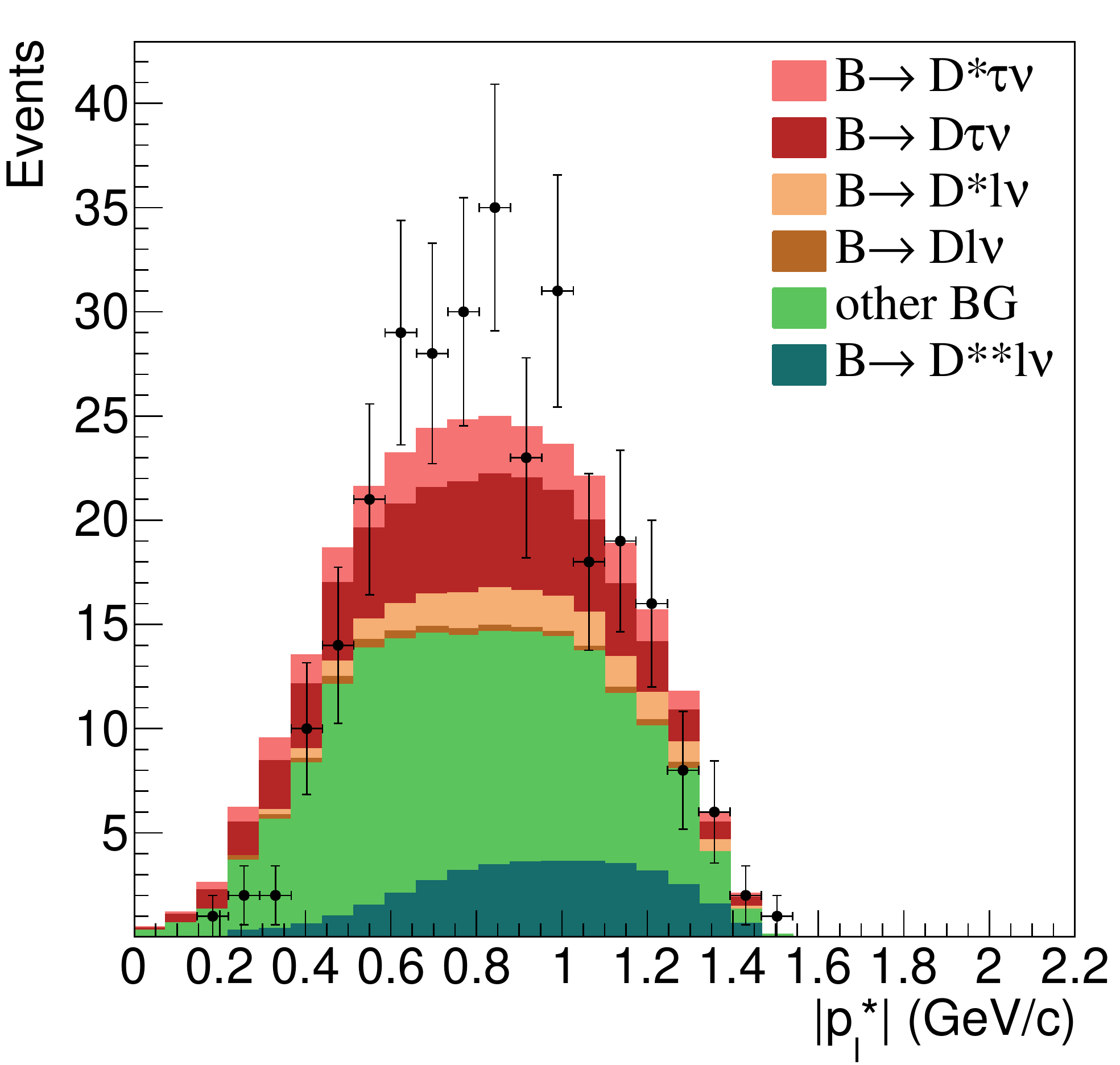}
	\includegraphics[width=\columnwidth]{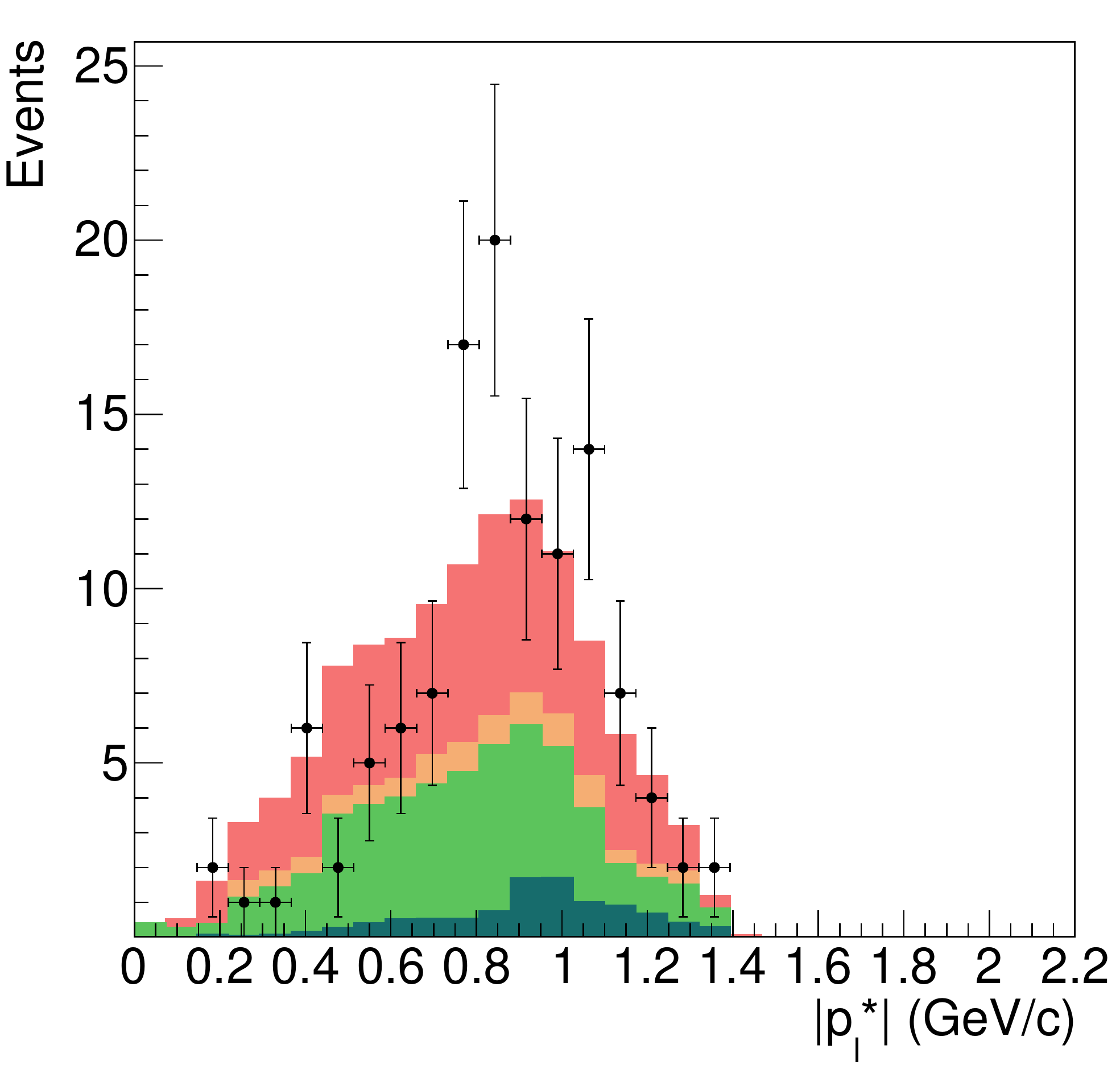}
	\includegraphics[width=\columnwidth]{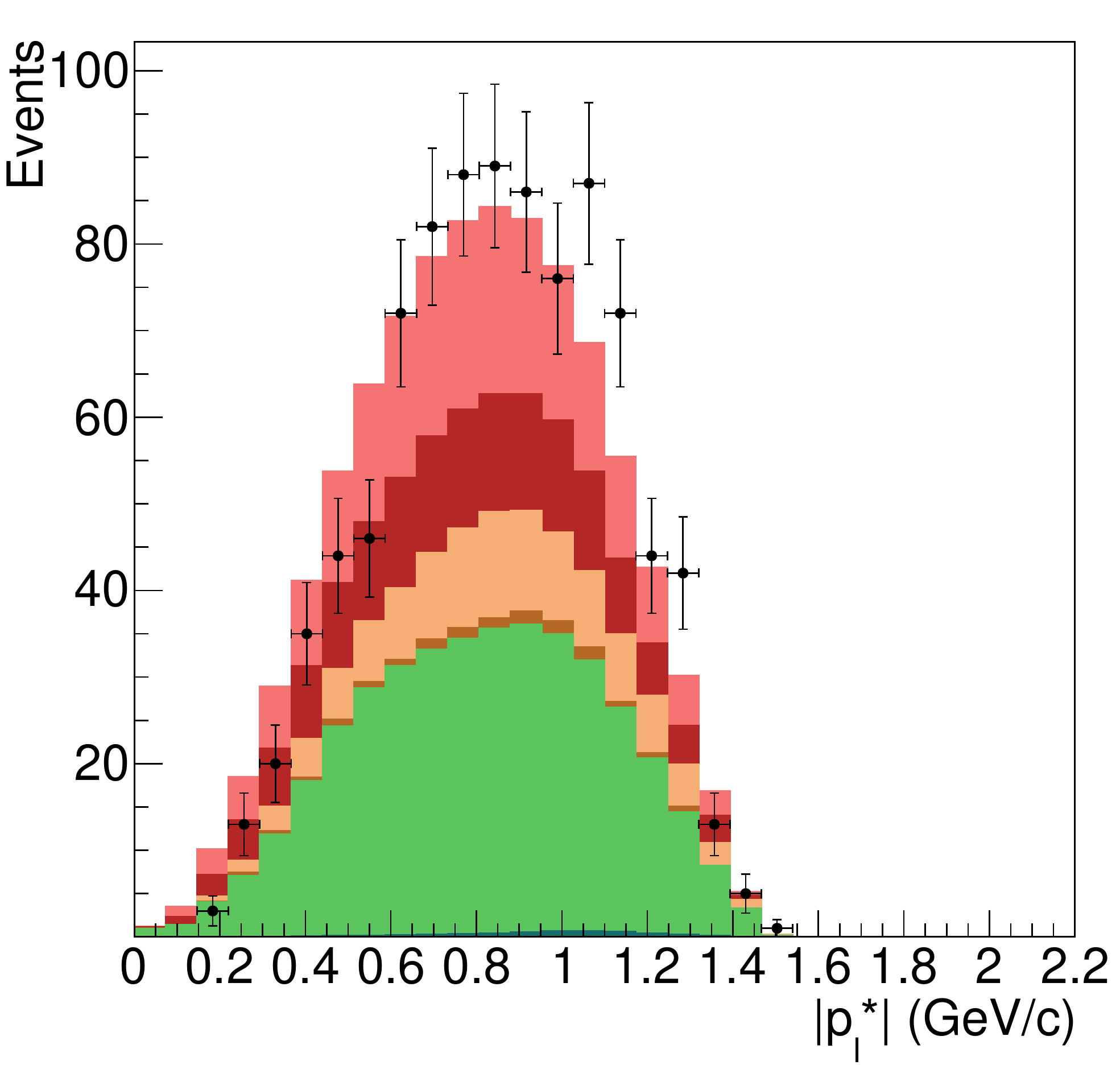}
	\includegraphics[width=\columnwidth]{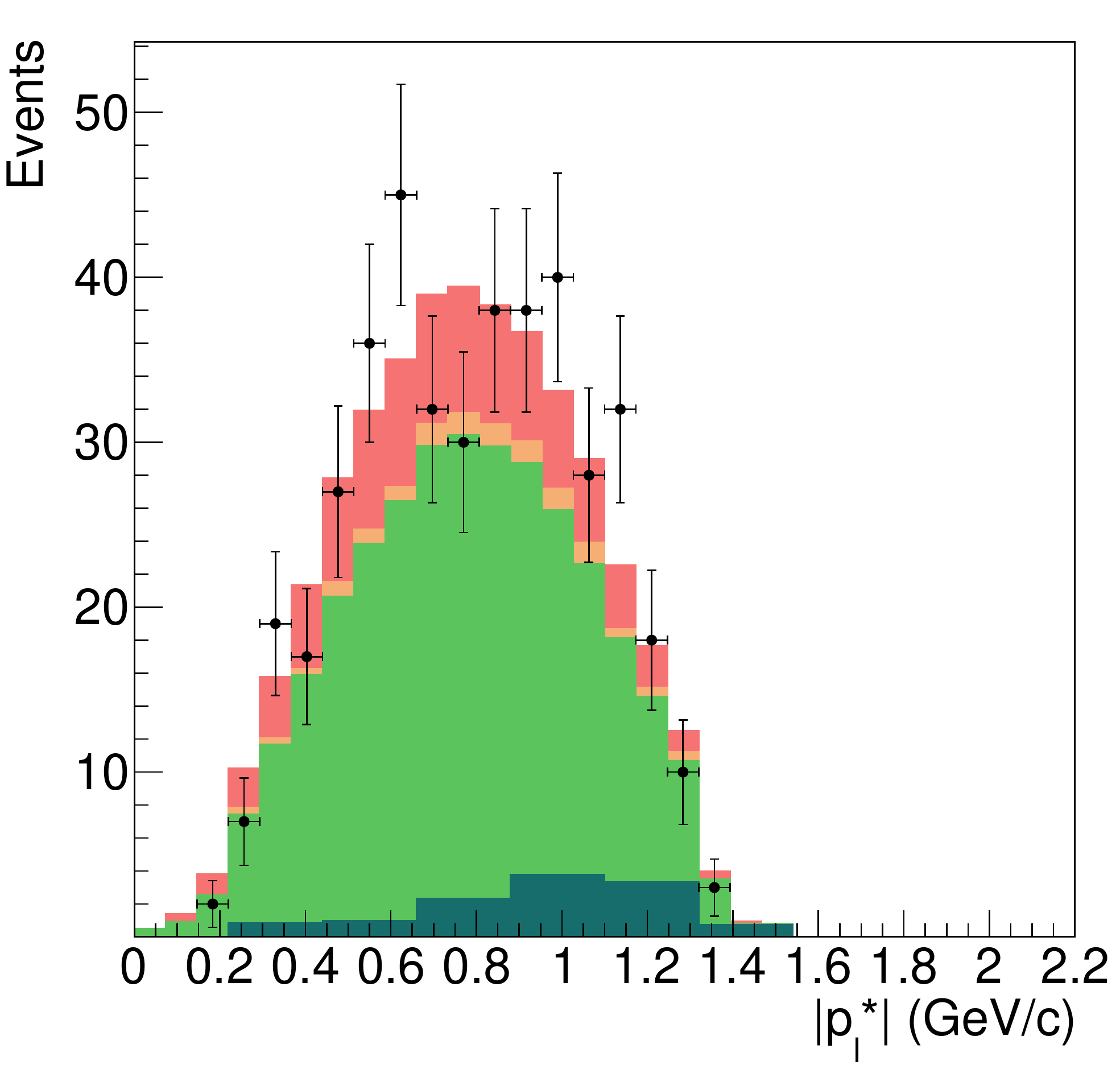}
\caption{Projections of the fit results and data points with statistical uncertainties in a signal-enhanced region of $\MM>\SI{2.0}{\unitmasssquared}$ in the $\PLEP$ dimension.
Top left: $D^{+}\ell^-$; top right: $D^{\ast +}\ell^-$; bottom left: $D^{0}\ell^-$; bottom right: $D^{\ast 0}\ell^-$.}
\label{fig:results:control_plep}
\end{figure*}

The fit to the entire data sample gives
\begin{eqnarray}
R(D) &=& 0.375 \pm 0.064 \\*
R(D^\ast) &=& 0.293 \pm 0.038 \ ,
\end{eqnarray}
corresponding to a yield of \num{320} $\BDdecay$ and \num{503} $\BDSdecay$ events; the errors are statistical.
Projections of the fit are shown in Figs.~\ref{fig:results:Dmodes} and \ref{fig:results:Dstmodes}.
The high-$\MM$ distributions and the fit projections are shown in Fig.~\ref{fig:results:control_mm}.
Figures~\ref{fig:results:control_eecl} and \ref{fig:results:control_plep} show the signal-enhanced ($\MM>\SI{2.0}{\unitmasssquared}$) fit projections in $\ecl$ (the most powerful classifier in the neural network) and $\PLEP$, respectively.
In these figures, all background components except \component{$D^{\ast\ast}$ background} are combined into the \component{other-BG} component for clarity.
The best-fit yields are given in Table~\ref{tab:results:fit}.
\begin{table}[htb]
\caption{Fit results and expected yields as derived from simulated data.}
\label{tab:results:fit}
\begin{ruledtabular}
\centering
\begin{tabular}{llrr}
Sample           &	Component                                  &   Yield & Expected yield \\ \hline
$D^+\ell^-$        &	$\ell$ normalization  &   $844 \pm 34$ & $870$ \\
$D^+\ell^-$        &	$\ell$ CF             &   $924 \pm 47$ & $970$ \\
$D^+\ell^-$        &	$D^{\ast\ast}$ BG     &   $108 \pm 38$ & $133$ \\
$D^0\ell^-$        &	$\ell$ normalization  &  $2303 \pm 64$ & $2290$ \\
$D^0\ell^-$        &	$\ell$ CF             &  $7324 \pm 122$ & $7440$ \\
$D^0\ell^-$        &	$D^{\ast\ast}$ BG     &   $131 \pm 81$ & $210$ \\
$D^{\ast +}\ell^-$ &	$\ell$ normalization  &  $1609 \pm 43$ & $1680$ \\
$D^{\ast +}\ell^-$ &	$D^{\ast\ast}$ BG     &    $36 \pm 18$ & $76$ \\
$D^{\ast 0}\ell^-$ &	$\ell$ normalization  &  $2188 \pm 60$ & $2280$ \\
$D^{\ast 0}\ell^-$ &	$D^{\ast\ast}$ BG     &   $117 \pm 39$ & $40$ \\
\end{tabular}
\end{ruledtabular}
\end{table}

From the fit, the correlation between $R$ and $R^\ast$ is \num{-0.56}; each, in turn,
is most strongly correlated with the \component{$D^{\ast\ast}$ background} yields, with $0.1$ to $0.2$ for $R$ and $\approx 0.3$ for $R^\ast$.

\section{Systematic Uncertainties}
The dominant systematic uncertainties arise from our limited understanding of the $D^{\ast\ast}$ background and from uncertainties in the fixed factors used in the fit.
They are summarized in Table~\ref{tab:systematics:summary} and itemized below.
\begin{table}[htb]
\caption{Overview of relative systematic uncertainties in percent. The last column gives the correlation between $R(D)$ and $R(D^\ast)$.}
\label{tab:systematics:summary}
\begin{ruledtabular}
\centering
\begin{tabular}{rrrr}
	                                             & $R(D)\,[\%]$ & $R(D^\ast)\,[\%]$ & Correlation \\ \hline
	            $D^{(\ast(\ast))}\ell\nu$ shapes &       4.2 &            1.5 &        0.04 \\
	                  $D^{\ast\ast}$ composition &       1.3 &            3.0 &       -0.63 \\
	                             Fake $D$ yield  &       0.5 &            0.3 &        0.13 \\
	                           Fake $\ell$ yield &       0.5 &            0.6 &       -0.66 \\
	                                 $D_s$ yield &       0.1 &            0.1 &       -0.85 \\
          	                          Rest yield &       0.1 &            0.0 &       -0.70 \\
	                                    Efficiency ratio $f^{D^+}$ &       2.5 &            0.7 &       -0.98 \\
	                                     Efficiency ratio $f^{D^0}$ &       1.8 &            0.4 &        0.86 \\
	                 Efficiency ratio $f^{D^{\ast +}}_\mathrm{eff}$ &       1.3 &            2.5 &       -0.99 \\
	                 Efficiency ratio $f^{D^{\ast 0}}_\mathrm{eff}$ &       0.7 &            1.1 &        0.94 \\
	                                      CF double ratio $g^+$ &       2.2 &            2.0 &       -1.00 \\
	                                      CF double ratio $g^0$ &       1.7 &            1.0 &       -1.00 \\
	                             Efficiency ratio $f_\mathrm{wc}$ &       0.0 &            0.0 &        0.84 \\
	                                 $\MM$ shape &       0.6 &            1.0 &        0.00 \\
	                               $\NBTR$ shape &       3.2 &            0.8 &        0.00 \\
	                       Lepton PID efficiency &       0.5 &            0.5 &        1.00 \\ \hline
	                                       Total &       7.1 &            5.2 &      $-0.32$
\end{tabular}
\end{ruledtabular}
\end{table}

In the table, ``$D^{(\ast(\ast))}\ell\nu$ shapes'' refers to uncertainties in the parameters that are used for the shape reweighting of semileptonic decays.
The effect on the result is extracted by creating different sets of weights according to shape hypotheses from varying individual production parameters within their $1\sigma$ limits.

The \component{$D^{\ast \ast}$ background} has a strong influence on the extracted yield of the \component{tau signal} because the two components overlap in the $\MM$ spectrum.
In addition to the shape uncertainties, there are uncertainties related to the poorly determined branching fractions to the different $D^{\ast\ast}$ states.
The fit is therefore repeated several times: twice for each $D^{\ast\ast}$ state, with its branching fractions varied within its uncertainties.
We use the following uncertainties: \SI{42.3}{\percent} for $D_2^\ast$, \SI{34.6}{\percent} for $D_0^\ast$, \SI{14.9}{\percent} for $D_1$, \SI{36.2}{\percent} for $D_1^\prime$, and \SI{100.0}{\percent} for the radially excited $D(2S)$ and $D^\ast(2S)$.
The best-fit variations in $R$ are used as systematic uncertainties.
They are combined quadratically and quoted in Table~\ref{tab:systematics:summary} as ``$D^{\ast\ast}$ composition.''

All fixed factors used in the fit are varied by their uncertainty (arising from the MC sample size).
The influence of the uncertainty of these factors is shown individually in Table~\ref{tab:systematics:summary}.
Most factors---especially the fixed yields---have little influence on the overall uncertainty;
the efficiency ratios $f^{D^{+,0}}$ and $f^{D^{\ast +,0}}_\mathrm{eff}$ and the cross-feed probability ratios $g^{+,0}$ give the largest contributions, comparable to the $D^{\ast\ast}$ composition and $D^{(\ast(\ast))}\ell\nu$ shape uncertainties.

To evaluate the effect of PDF uncertainties, the shapes of all components are modified and the fit is repeated. 
The nominal fit uses smoothed-histogram PDFs in $\MM$; here, these are replaced by unsmoothed-histogram PDFs.
The variation of the best-fit $R$ is taken as the symmetric systematic uncertainty for ``$\MM$ shape'' in Table~\ref{tab:systematics:summary}.
For the $\NBTR$ alternate model, we replace the bifurcated Gaussians by kernel-estimator functions with adaptive bandwidth.
Again, the deviation from the nominal fit value is taken as the symmetric systematic uncertainty for ``$\NBTR$ shape'' in Table~\ref{tab:systematics:summary}.
It is among the dominant systematic uncertainties.

The identification efficiencies for primary and secondary leptons are slightly different between simulated and real data.
This difference affects the measurement by modifying the efficiency ratios.
It has been calibrated for different lepton kinematics and run conditions using $J/\psi\to\ell^+\ell^-$ decays,
leading to a \SI{0.5}{\percent} relative uncertainty in $R(D)$ and $R(D^\ast)$.

The correlations of $R(D)$ and $R(D^\ast)$ for each itemized systematic-uncertainty contribution are given in the last column of Table~\ref{tab:systematics:summary}.
These are calculated using 500 pseudoexperiments, with two exceptions:
the shape uncertainties are assumed to be uncorrelated while the lepton ID efficiencies are assumed to be $100\%$ correlated between $R(D)$ and $R(D^\ast)$.
The total correlation of the systematic uncertainties is \num{-0.32}.

\section{Results and Discussion}
The best-fit results, including systematic uncertainties, are
\begin{eqnarray}
R(D) &=& 0.375 \pm 0.064 \pm 0.026 \\*
R(D^\ast) &=& 0.293 \pm 0.038 \pm 0.015 \ .
\end{eqnarray}

\begin{figure}[htb]
\includegraphics[width=\columnwidth]{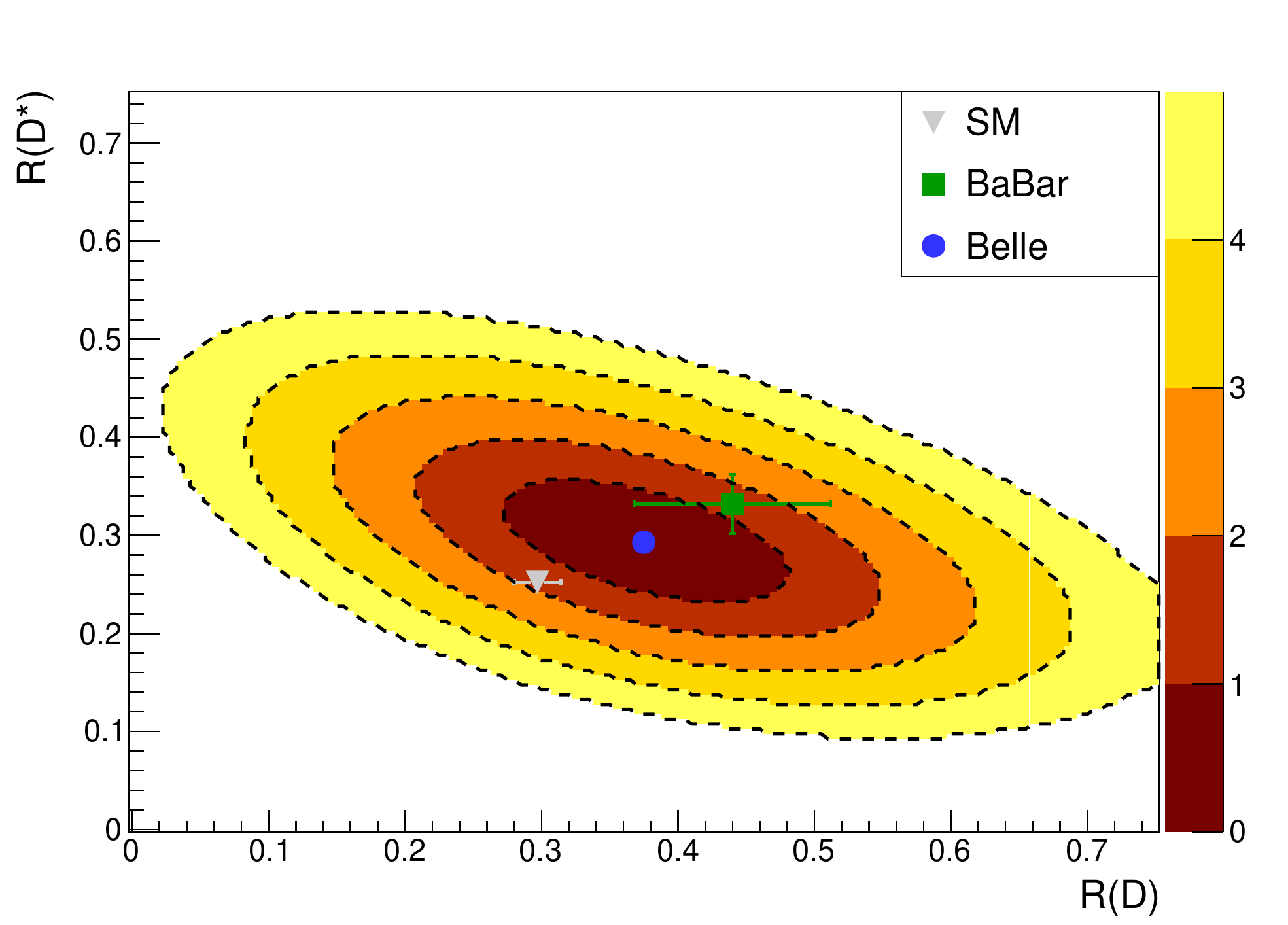}
\caption{Exclusion level of $R(D)$-$R(D^\ast)$ value assumptions in standard deviations, systematic uncertainties included.}
\label{fig:discussion:rrstplane}
\end{figure}

Figure~\ref{fig:discussion:rrstplane} shows the exclusion level in the $R(D)$--$R(D^\ast)$ plane, based on the likelihood distribution that is convoluted with a correlated two-dimensional normal distribution according to the systematic uncertainties.
The exclusions of the central values of the BaBar measurement~\cite{Lees:2012xj} and the SM prediction as determined in Ref.~\cite{Lees:2012xj} are comparably low at $1.4\sigma$ and $1.8\sigma$, respectively.
While our measurement does not favor one over the other, both measurements deviate in the same direction from the SM expectation.

\begin{figure}[htb]
\includegraphics[width=\columnwidth]{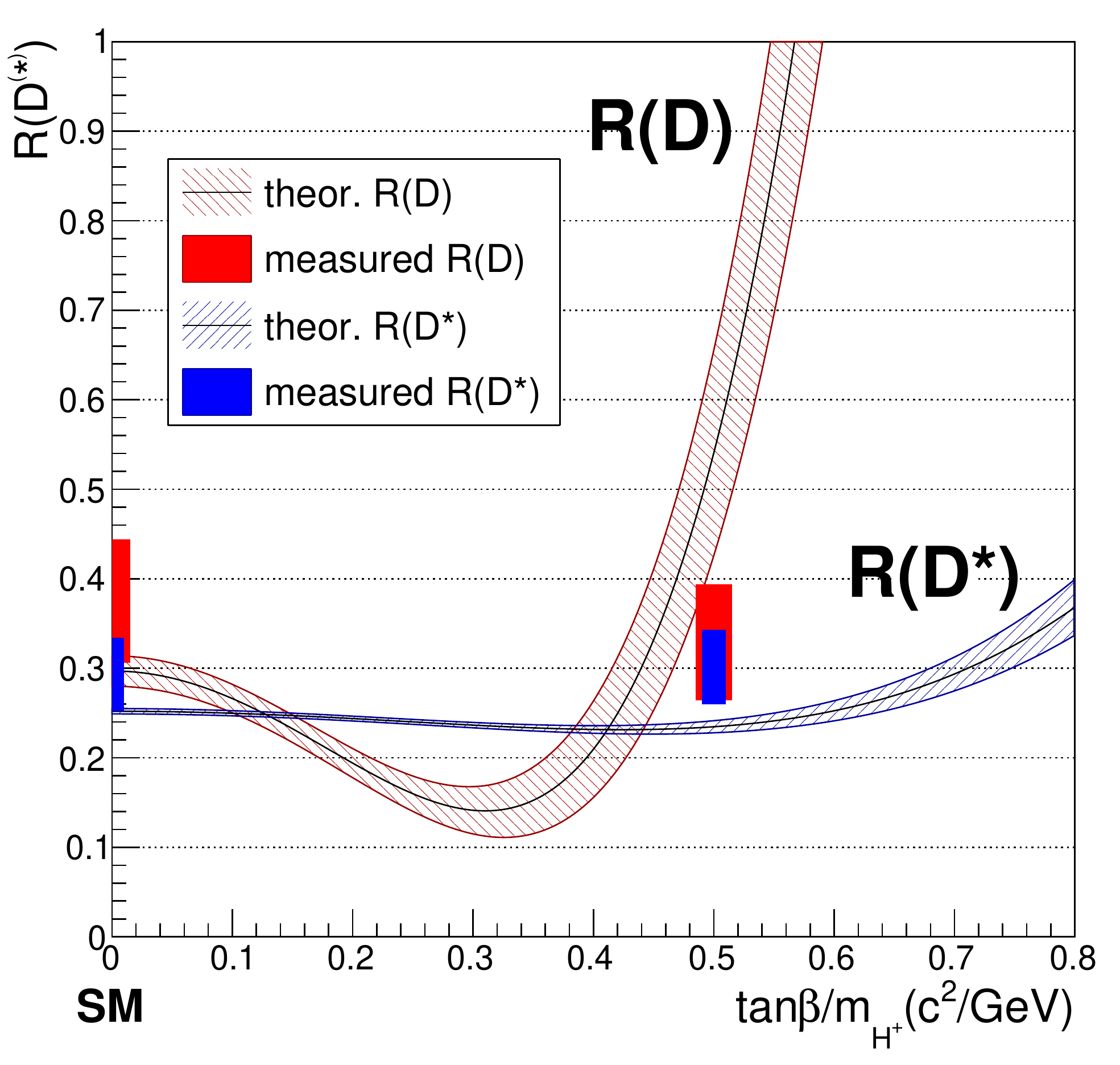}
\caption{Theoretical predictions with \SI{1}{\stddev} error ranges for $R(D)$ (red) and $R(D^\ast)$ (blue) for different values of $\tan{\beta}/m_{H^+}$ in the 2HDM of type II. The fit results for $\tan{\beta}/m_{H^+}=\SI{0.5}{\clight\squared/\giga\electronvolt}$ and SM are shown with their \SI{1}{\stddev} ranges as red and blue bars with arbitrary width for better visibility.}
\label{fig:discussion:rrst_1d}
\end{figure}

We also use our fit procedure to test the compatibility of the data samples with the two-Higgs-doublet model of type II.
For this purpose, we perform the analysis with the 2HDM MC sample with $\tan{\beta}/m_{H^+}=\SI{0.5}{\clight\squared/\giga\electronvolt}$ to extract probability density distributions.
The best-fit values in this alternate model are
\begin{eqnarray}
R(D) &=& 0.329 \pm 0.060\mathrm{(stat.)} \pm 0.022\mathrm{(syst.)}\\*
R(D^\ast) &=& 0.301 \pm 0.039\mathrm{(stat.)} \pm 0.015\mathrm{(syst.)}\ .
\end{eqnarray}
The effect on the measured $R(D^\ast)$ value is very small but the measured value for $R(D)$ is significantly lower.
For the prediction in the 2HDM of type II, we use formula~(20) in Ref.~\cite{Lees:2012xj};
the expected values are
\begin{eqnarray}
R(D)_\mathrm{2HDM} &=& 0.590\pm0.125\\*
R(D^\ast)_\mathrm{2HDM} &=& 0.241\pm0.007\ .
\end{eqnarray}
Figure~\ref{fig:discussion:rrst_1d} shows the predictions of $R(D)$ and $R(D^\ast)$ as a function of $\tan{\beta}/m_{H^+}$ for the type II 2HDM, together with our results for the two studied values of $0$ (SM) and $\SI{0.5}{\clight\squared/\giga\electronvolt}$.
In contrast to BaBar's measurements, our results are compatible with the type II 2DHM in the $\tan{\beta}/m_{H^+}$ regions around $\SI{0.45}{\clight\squared/\giga\electronvolt}$ and zero.

\begin{figure*}
	\includegraphics[width=\columnwidth]{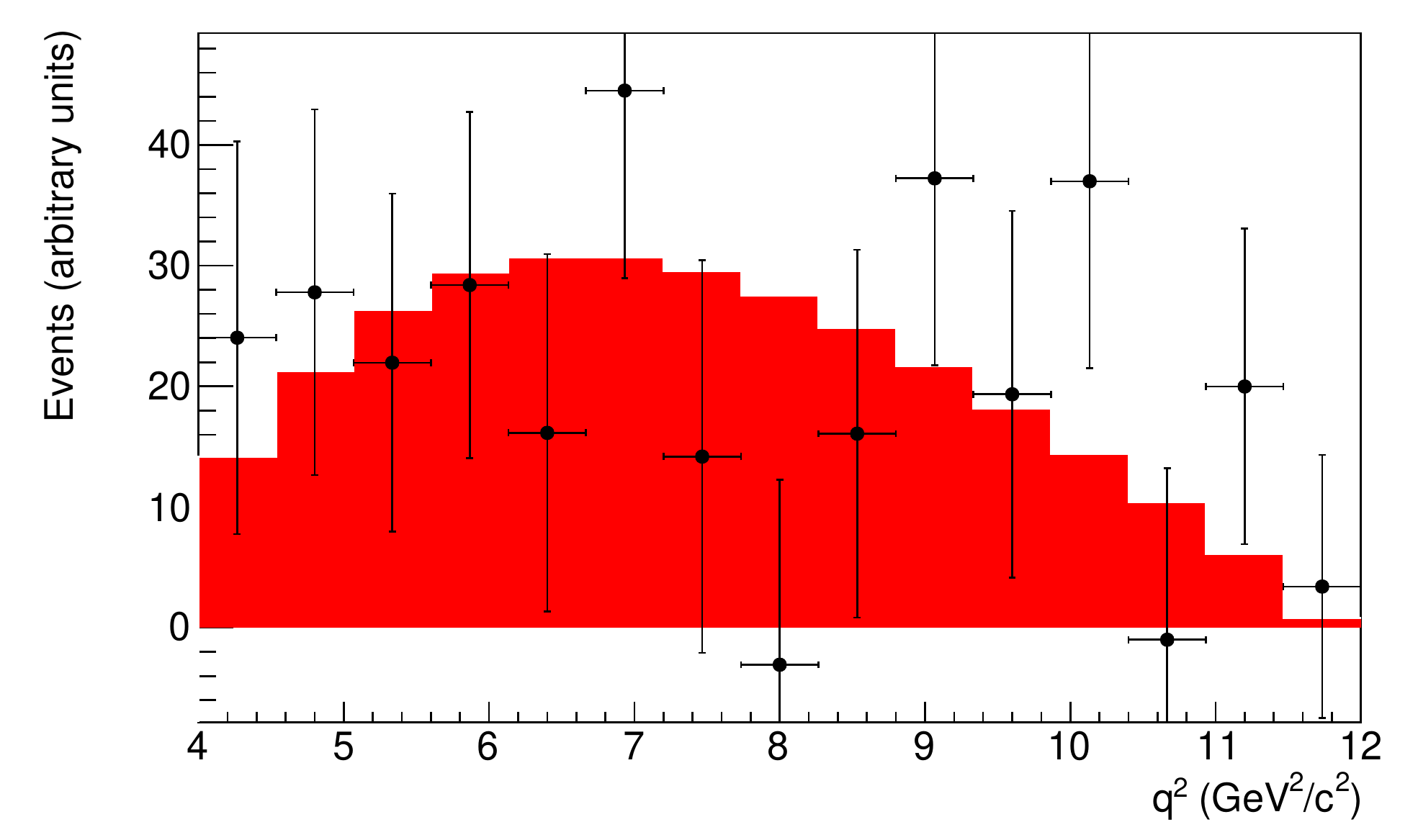}
	\includegraphics[width=\columnwidth]{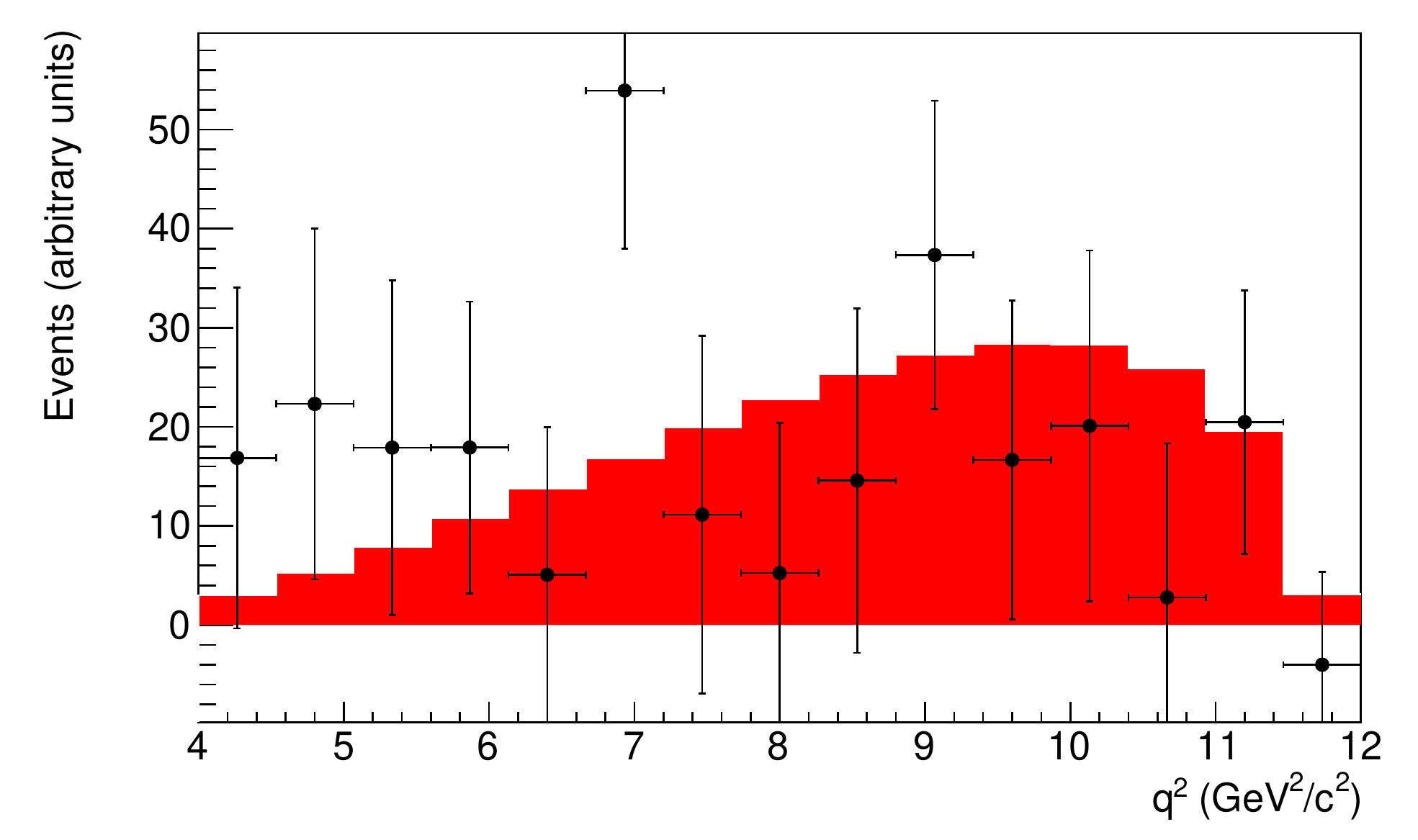}
	\includegraphics[width=\columnwidth]{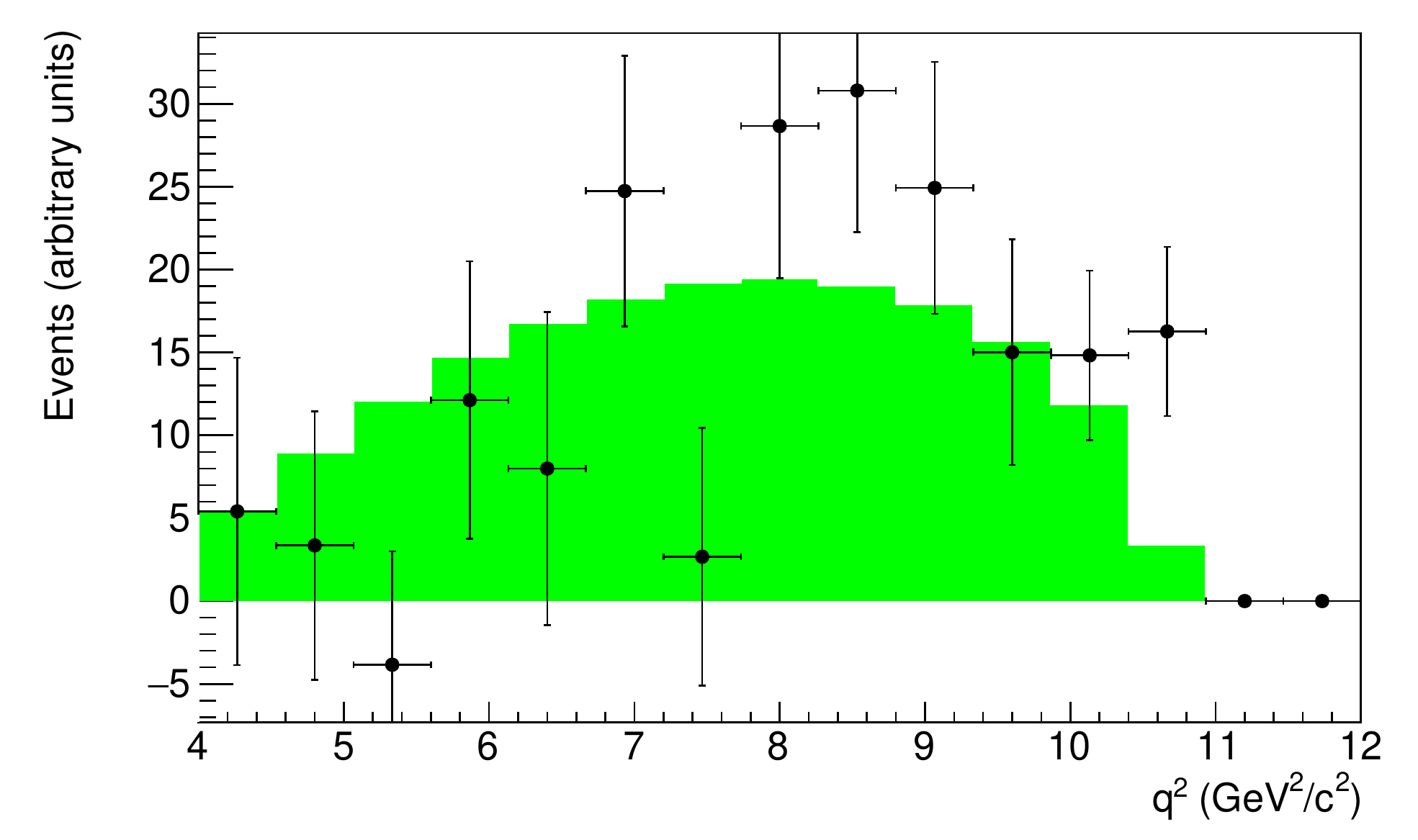}
	\includegraphics[width=\columnwidth]{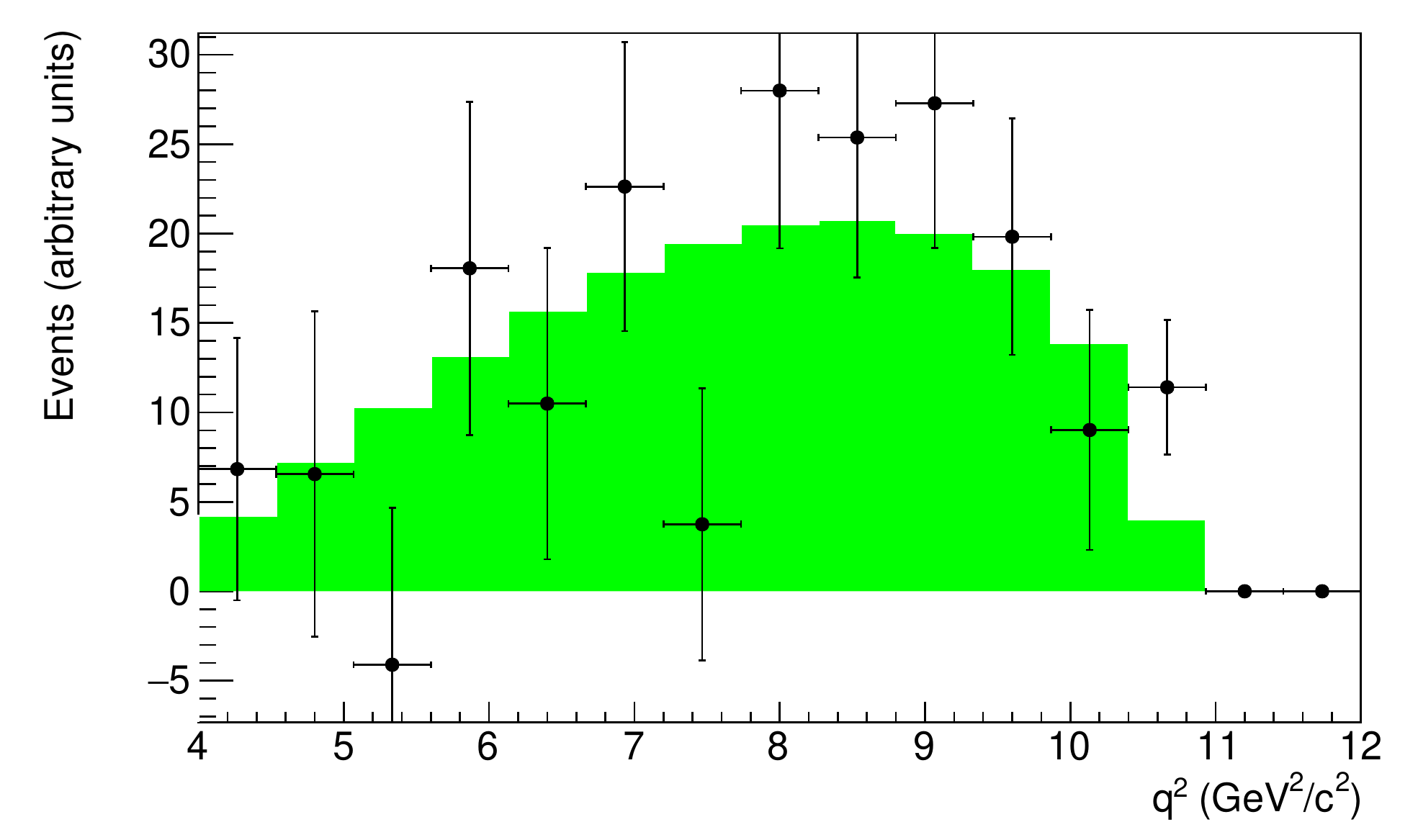}
\caption{Background-subtracted $\QSQ$ distributions of the $\tau$ signal in the region of $\MM>\SI{0.85}{\unitmasssquared}$. The distributions are efficiency corrected and normalized to the fitted yield. The error bars show the statistical uncertainties. 
The histogram is the respective expected distribution from signal MC. Left: Standard Model result, right: Type-II 2HDM result with $\tan{\beta}/m_{H^+}=\SI{0.5}{\clight\squared/\giga\electronvolt}$, top: $\BDdecay$, bottom: $\BDSdecay$}
\label{fig:discussion:np:q2}
\end{figure*}

The observable most sensitive to NP extensions of the SM with a scalar charged Higgs is $\QSQ$.
We estimate the signal $\QSQ$ distributions by subtracting the background, using the distributions from simulated data and the yields from the fit procedure, and correcting the distributions using efficiency estimations from simulated data.
The $D^+\ell^-$ and $D^0\ell^-$ samples and the $D^{\ast+}\ell^-$ and $D^{\ast0}\ell^-$ samples are combined to increase the available statistics, then the full procedure is repeated using the assumptions for the \component{$\tau$ signal} in a type II 2HDM model with $\tan{\beta}/m_{H^+}=\SI{0.5}{\clight\squared/\giga\electronvolt}$.
Figure~\ref{fig:discussion:np:q2} shows the measured background-subtracted and efficiency-corrected $\QSQ$ distributions for the SM and the NP point.
As the signal yields are not extracted from fits to individual $\QSQ$ bins, the data distribution depends slightly on the signal model; the signal model can affect the background yields in the fit to uncorrected data, which are then subtracted.
A $\chi^2$ test shows that both hypotheses are compatible with our data with $p$-values for the SM distribution of $64\%$ ($D\tau^-\bar{\nu}_\tau$) and $11\%$ ($D^\ast\tau^-\bar{\nu}_\tau$), and for the NP distribution of $53\%$ ($D\tau^-\bar{\nu}_\tau$) and $49\%$ ($D^\ast\tau^-\bar{\nu}_\tau$).

\section{Conclusion}
We present a measurement of the relative branching ratios $R(D^{(\ast)})$ of $\Bdecay$ to $\Bdecaynorm$ using the full $\Upsilon(4S)$ data recorded with the Belle detector. The results are
\begin{eqnarray*}
R(D) &=& 0.375 \pm 0.064\mathrm{(stat.)}\pm 0.026\mathrm{(syst.)}\\*
R(D^\ast) &=& 0.293 \pm 0.038\mathrm{(stat.)}\pm 0.015\mathrm{(syst.)}\ .
\end{eqnarray*}
In comparison to our previous preliminary results~\cite{Adachi:2009qg}, which are superseded by this measurement, we utilize a more sophisticated fit strategy with an improved handling of the background from $\bar{B}\to D^{\ast\ast}\ell^-\bar{\nu}_\ell$ events, impose an isospin constraint, and exploit a much higher tagging efficiency. 
By these methods, we reduce the statistical uncertainties by about a third and the systematic uncertainties by more than a half.

Our result lies between the SM expectation and the most recent measurement from the BaBar collaboration~\cite{Lees:2012xj} and is compatible with both.
It is also compatible with a 2HDM of type II in the region around $\tan{\beta}/m_{H^+}=\SI{0.5}{\clight\squared/\giga\electronvolt}$, as illustrated in Figs.~\ref{fig:discussion:rrst_1d} and \ref{fig:discussion:np:q2}.

\section{Acknowledgments}
\begin{acknowledgments}
We thank the KEKB group for the excellent operation of the
accelerator; the KEK cryogenics group for the efficient
operation of the solenoid; and the KEK computer group,
the National Institute of Informatics, and the 
PNNL/EMSL computing group for valuable computing
and SINET4 network support.  We acknowledge support from
the Ministry of Education, Culture, Sports, Science, and
Technology (MEXT) of Japan, the Japan Society for the 
Promotion of Science (JSPS), and the Tau-Lepton Physics 
Research Center of Nagoya University; 
the Australian Research Council and the Australian 
Department of Industry, Innovation, Science and Research;
Austrian Science Fund under Grants No.~P 22742-N16 and P 26794-N20;
the National Natural Science Foundation of China under Contracts 
No.~10575109, No.~10775142, No.~10875115, No.~11175187, and  No.~11475187; 
the Ministry of Education, Youth and Sports of the Czech
Republic under Contract No.~LG14034;
the Carl Zeiss Foundation, the Deutsche Forschungsgemeinschaft
and the VolkswagenStiftung;
the Department of Science and Technology of India; 
the Istituto Nazionale di Fisica Nucleare of Italy; 
National Research Foundation (NRF) of Korea Grants
No.~2011-0029457, No.~2012-0008143, No.~2012R1A1A2008330, 
No.~2013R1A1A3007772, No.~2014R1A2A2A01005286, No.~2014R1A2A2A01002734, 
No.~2014R1A1A2006456;
the Basic Research Lab program under NRF Grants No.~KRF-2011-0020333, 
No.~KRF-2011-0021196, Center for Korean J-PARC Users, No.~NRF-2013K1A3A7A06056592; 
the Brain Korea 21-Plus program and the Global Science Experimental Data 
Hub Center of the Korea Institute of Science and Technology Information;
the Polish Ministry of Science and Higher Education and 
the National Science Center;
the Ministry of Education and Science of the Russian Federation and
the Russian Foundation for Basic Research;
the Slovenian Research Agency;
the Basque Foundation for Science (IKERBASQUE) and 
the Euskal Herriko Unibertsitatea (UPV/EHU) under program UFI 11/55 (Spain);
the Swiss National Science Foundation; the National Science Council
and the Ministry of Education of Taiwan; and the U.S.\
Department of Energy and the National Science Foundation.
This work is supported by a Grant-in-Aid from MEXT for 
Science Research in a Priority Area (New Development of 
Flavor Physics) and from JSPS for Creative Scientific 
Research (Evolution of Tau-lepton Physics).
\end{acknowledgments}


\begin{thebibliography}{99}
\bibitem{CC}
Throughout this paper, 
the inclusion of the charge-conjugate decay mode is implied.
\end{thebibliography}
\end{document}